\documentclass[aps,pra,nofootinbib,floatfix,preprintnumbers,superscriptaddress,showkeys,12pt]{revtex4-2}
\pdfoutput=1

\usepackage{amssymb,amsfonts,amsthm,amsmath}
\usepackage{mathtools}
\usepackage{verbatim}
\usepackage{theoremref}
\usepackage{mathrsfs}
\RequirePackage[OT1]{fontenc}

\usepackage{bm}
\usepackage[dvipsnames]{xcolor}
\usepackage{titlesec}
\usepackage{textcomp}
\usepackage{fullpage}
\usepackage{bbm}
\usepackage[section]{placeins}
\usepackage{comment}
\usepackage{accents}
\usepackage{appendix}

\usepackage{tikz}
\usetikzlibrary{positioning}
\usetikzlibrary{calc}
\pgfdeclarelayer{bg}    
\pgfsetlayers{bg,main}  
\usepackage{graphicx}
\usepackage{caption}
\usepackage{subcaption}
\usepackage{float}
\graphicspath{ {./figures/} }

\usepackage{math-macros}

\begin{document}

\title{Quantum Distance Approximation for Persistence Diagrams}

\author{Bernardo Ameneyro}
\email{bameneyr@vols.utk.edu}
\affiliation{Department of Mathematics, The University of Tennessee, Knoxville, TN 37996-1320, USA }

\author{Rebekah Herrman}
\email{rherrma2@utk.edu}
\affiliation{Department of Industrial and Systems Engineering, The University of Tennessee, Knoxville, TN 37996, USA}

\author{George Siopsis}
\email{siopsis@tennessee.edu}
\affiliation{Department of Physics and Astronomy, The University of Tennessee, Knoxville, TN 37996-1200, USA}

\author{Vasileios Maroulas}
\email{vasileios.maroulas@utk.edu}
\affiliation{Department of Mathematics, The University of Tennessee, Knoxville, TN 37996-1320, USA }

\begin{abstract}
    Topological Data Analysis methods can be useful for classification and clustering tasks in many different fields as they can provide two dimensional persistence diagrams that summarize important information about the shape of potentially complex and high dimensional data sets.
    The space of persistence diagrams can be endowed with various metrics such as the Wasserstein distance which admit a statistical structure and allow to use these summaries for machine learning algorithms.
    However, computing the distance between two persistence diagrams involves finding an optimal way to match the points of the two diagrams and may not always be an easy task for classical computers.
    In this work we explore the potential of quantum computers to estimate the distance between persistence diagrams, in particular we propose variational quantum algorithms for the Wasserstein distance as well as the \(d^{c}_{p}\) distance.
    Our implementation is a weighted version of the Quantum Approximate Optimization Algorithm that relies on control clauses to encode the constraints of the optimization problem.
\end{abstract}

\keywords{topological data analysis, distances of persistence diagrams, quantum approximate optimization algorithms}

\maketitle


\section{Introduction} \label{sec:intro}
Topological data analysis (TDA) methods aim to characterize data shape using topological properties, which are invariant under continuous transformations like rotations and scaling.
TDA approaches can reduce the dimensionality of large and complex data sets, and are often robust to noise.
Naturally, they have found numerous applications to different fields, from biology and healthcare \cite{chung2015, tda_wheeze, tda_clustering2015, gunnarcancer, HodgeCycle, CPD_me, Maroulas2021, jholes, persis_brain}, to chemistry and materials science \cite{Maroulas2020, Townsend2020, NA2022, Papamarkou2022, Chen2021, fullerene}, image classification for action recognition, robotics and handwriting analysis \cite{tda_action, vasudevan2013, tda_number}, and even sensor networks \cite{GdS06, GdS07, Ghrist12, CdS1, CdS2, persissensor}.
In addition, TDA techniques can be used to study complex dynamical systems and identify properties such as multi-stability, periodicity and chaos.
So, TDA approaches for time series data sets exist \cite{tda_timeseries, tda_windowgenes} and are used to solve problems like signal analysis and identification \cite{tda_windows, tda_signal, Marchese2016, Marchese2018}, often capturing results that other traditional signal analysis methods miss.
With the recent rise of quantum computers, a few works have appeared exploring the potential advantages they can provide for TDA methods \cite{ameneyro2024quantum, lloyd2016quantum, siopsis2019quantum, hayakawa2022quantum, ameneyro2022takens}.

TDA methods first extract topological features from the data -- such as the number of connected components, holes, and voids -- via persistent homology, tracking them across different scales or resolutions \cite{zomo-tda, CompyTopo, carlsson2021topological}.
The topological features are then displayed in persistence diagrams that show when each feature appears and disappears.
These 2-dimensional diagrams can summarize large and high-dimensional data sets in order to perform machine learning algorithms for classification and clustering, but to do this one must properly define and compute distances on the space of persistence diagrams.
Although there are various distances, the focus in this paper is the consideration of the popular  Wasserstein distance, and the one defined in \cite{Marchese2018}, which has been proven to be advantageous in machine learning tasks, e.g., see \cite{kmeans}.
However, computing any distance between two persistence diagrams involves minimizing a cost function over all the possible ways to match the points in the diagrams, which can be a challenging process for classical computers if the diagrams have a large number of points.

The first quantum approach to compute the distance between two persistence diagrams formulates the Wasserstein distance as a quadratic unconstrained binary optimization problem (QUBO) and uses quantum annealing through a D-Wave 2000Q quantum computer to estimate the solution \cite{quantumWasserstein}. 
Quantum Approximate Optimization Algorithms (QAOA) and other quantum variational algorithms have been used to solve similar combinatorics problems such as sampling from maximal matchings of graphs \cite{qaoaGraphMatching} and the MaxCut problem \cite{qaoaWeights}.
More recently, Saravanan and Gopikrishnan introduced in \cite{qaoaWasserstein} a QAOA approach to compute the Wasserstein distance between persistence diagrams.
Their work also formulates a QUBO problem and suggests the use of a QAOA algorithm to estimate the solution, but it lacks details on the complexity and performance of the algorithm.

In this paper, we propose a QAOA inspired approach to compute the distance between persistence diagrams using a quantum computer. 
Indeed, we provide a fully quantum algorithm to estimate both the Wassertsein distance and the one proposed in \cite{Marchese2018} between persistence diagrams.
We formulate these distances as constrained optimization problems, and then use a Hamiltonian to encode the cost of the optimization problem and control clauses \cite{qaoaGraphMatching} to implement the constraints.
We build a mixing operator that produces all the quantum states of interest for the optimization in just one iteration.
Furthermore, the overall complexity of the quantum algorithm is \(\mathcal{O}(mn)\), where \(n \le m\) are the cardinalities of the persistence diagrams, in contrast to the Hungarian algorithm used to compute the distances classically with a cost of \(\mathcal{O}(m n^{2})\).

This paper is organized as follows.
First, Section \ref{sec:background} provides background information on TDA and QAOA.
Then, Section \ref{sec:methods} details the optimization problems as well as the quantum algorithms to solve them.
Next, Section \ref{sec:results} shows simple illustrative examples and simulations of the quantum algorithms.
Last, Section \ref{sec:conclusion} concludes with a brief discussion of our results.
In addition, Appendix \ref{sec:proofs} contains proofs of some of the Theorems presented in Section \ref{sec:dcp}.

\section{Preliminary Results}\label{sec:background}

In what follows, we briefly introduce TDA and quantum information concepts that are relevant to our algorithm.
Any reader who wishes for more details on the subject may consult \cite{carlsson, CompyTopo, carlsson2021topological} for references on persistent homology and \cite{qaoa, qaoaGraphMatching, qaoaWeights, herrman2022multi} for QAOA references.

\subsection{Persistent Homology}

\begin{figure}
\centering
\begin{tikzpicture}[line join = round, line cap = round]

  \coordinate (v0) at (-4, 0, 0);

  \coordinate (e1) at (-2,0,1);
  \coordinate (e0) at ({-2+.5*sqrt(3)},0,-.5);

  \coordinate (t2) at (1,{sqrt(2)},0);
  \coordinate (t1) at (1,0,1);
  \coordinate (t0) at ({1+.5*sqrt(3)},0,-.5);

  \coordinate (th3) at (4,{sqrt(2)},0);
  \coordinate (th2) at ({4-.5*sqrt(3)},0,-.5);
  \coordinate (th1) at (4,0,1);
  \coordinate (th0) at ({4+.5*sqrt(3)},0,-.5);

  \begin{scope}
    \fill[RoyalBlue!50, draw=Blue, thick] (v0) circle (3pt) ;

    \fill[RoyalBlue!50, draw=Blue, thick] (e0) circle (3pt) ;
    \fill[RoyalBlue!50, draw=Blue, thick] (e1) circle (3pt) ;

    \fill[RoyalBlue!50, draw=Blue, thick] (t0) circle (3pt) ;
    \fill[RoyalBlue!50, draw=Blue, thick] (t1) circle (3pt) ;
    \fill[RoyalBlue!50, draw=Blue, thick] (t2) circle (3pt) ;

    \fill[RoyalBlue!50, draw=Blue, thick] (th0) circle (3pt) ;
    \fill[RoyalBlue!50, draw=Blue, thick] (th1) circle (3pt) ;
    \fill[RoyalBlue!50, draw=Blue, thick] (th2) circle (3pt) ;
    \fill[RoyalBlue!50, draw=Blue, thick] (th3) circle (3pt) ;
  \end{scope}
  \begin{pgfonlayer}{bg}
    \draw[RoyalBlue, ultra thick] (e1)--(e0);
    
    \draw[fill=Orange,fill opacity=.5] (t1)--(t0)--(t2)--cycle;
    \draw[RoyalBlue, ultra thick] (t1)--(t0);
    \draw[RoyalBlue, ultra thick] (t1)--(t2);
    \draw[RoyalBlue, ultra thick] (t0)--(t2);
    
    \draw[RoyalBlue, ultra thick, densely dotted] (th0)--(th2);
    \draw[fill=Orange,fill opacity=.5] (th1)--(th0)--(th3)--cycle;
    \draw[fill=RubineRed,fill opacity=.5] (th2)--(th1)--(th3)--cycle;
    \draw[RoyalBlue, ultra thick] (th1)--(th0);
    \draw[RoyalBlue, ultra thick] (th1)--(th2);
    \draw[RoyalBlue, ultra thick] (th2)--(th3);
    \draw[RoyalBlue, ultra thick] (th1)--(th3);
    \draw[RoyalBlue, ultra thick] (th0)--(th3);
  \end{pgfonlayer}
\end{tikzpicture}
\caption{
Examples of simplices.
A vertex for dimension zero, a line for dimension one, a triangle for dimension two, and a tetrahedron for dimension three.}
\label{fig:simplices}
\end{figure}
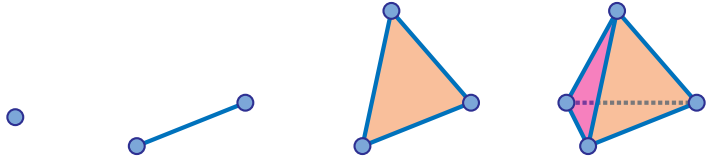

Homology is a mathematical field that aims to classify objects based on topological properties like connected components, holes, voids, and \(k\)-dimensional holes in general.
Its building blocks, simplices, are akin to graphs with higher order relationships and are written as an ordered list of their vertices \(\sigma = [v_{0}, \dots, v_{k}]\) (see Figure \ref{fig:simplices} for examples).
In turn, a simplicial complex \(\mathcal{K}\) is a collection of simplices satisfying that any face \(\tau \subseteq \sigma\) of a simplex \(\sigma\) in the complex, as well as the intersection \(\sigma_{1} \cap \sigma_{2}\) of two such simplices, also belong to the collection.
To illustrate this concept consider a point cloud data set in a metric space and fix a positive number \(\epsilon\), then the collection of simplices with diameter at most \(\epsilon\) using the data points as vertices is a simplicial (Vietoris-Rips) complex \(S^{\epsilon}\).
Given a simplicial complex \(\mathcal{K}\) one can define its chain group of dimension \(k\) as \(C_{k}(\mathcal{K}) = \left\{ \sum_{\sigma \in \mathcal{K}_{k}} \alpha_{\sigma} \sigma : \alpha_{\sigma} \in \mathbb{C} \right\}\), where \(\mathcal{K}_{k}\) are the simplices of dimension \(k\) in \(\mathcal{K}\) (those with \(k+1\) vertices).
Here we choose \(\mathbb{C}\) as the field since quantum systems are represented mathematically as Hilbert spaces over \(\mathbb{C}\).
Boundaries are fundamental to characterize topological properties like holes, so the concept is encoded as a homomorphism on chain groups.
In particular, the \(k\)-th boundary map \(\partial_{k} : C_{k}(\mathcal{K}) \rightarrow C_{k-1}(\mathcal{K})\) takes a simplex \(\sigma_{k} = [v_{0}, \dots, v_{k}] \in \mathcal{K}_{k}\) and returns the alternating sum \(\sum_{i=0}^{k} (-1)^{i} \hat{\sigma}_{k}(i)\) of its faces \(\hat{\sigma}_{k}(i) = [v_{0}, \dots, v_{i-1}, v_{i+1}, \dots, v_{k}]\) of dimension \(k-1\).
These boundary maps induce a chain complex \(\dots \xrightarrow{\partial_{k+1}} C_{k}(\mathcal{K}) \xrightarrow{\partial_{k}} C_{k-1}(\mathcal{K}) \xrightarrow{\partial_{k-1}} \dots \xrightarrow{\partial_{1}} C_{0}(\mathcal{K}) \xrightarrow{\bm 0} 0\) that satisfies \(\im{\partial_{k+1}} \subseteq \Ker{\partial_{k}}\) (see \cite{CompyTopo}).
Hence, we may define the \(k\)th homology group of \(\mathcal{K}\) as \(H_{k}(\mathcal{K}) = \Ker{\partial_{k}}/\im{\partial_{k+1}}\), which is generated by the \(k\)-dimensional holes of \(\mathcal{K}\).

Persistent homology extends this concept to consider sequences or filtrations of simplicial complexes \(\mathcal{K}_{1} \subseteq \dots \subseteq \mathcal{K}_{n}\) and track topological features across said sequences.
For instance, a Vietoris-Rips filtration \(S^{\epsilon_{1}} \subseteq \dots \subseteq S^{\epsilon_{n}}\) can be obtained by choosing a sequence of numbers \(0 \leq \epsilon_{1} \leq \dots \leq \epsilon_{n}\) representing different scales or resolutions in the data and building their respective Vietoris-Rips complexes \(S^{\epsilon_{i}}\).
The inclusion maps \(\iota : \mathcal{K}_{i} \rightarrow \mathcal{K}_{j}\) between nested complexes \(\mathcal{K}_{i} \subseteq \mathcal{K}_{j}\) induce group homomorphisms between the corresponding chain groups \(h_{k}^{i,j} = \iota_{*} : C_{k}(\mathcal{K}_{i}) \rightarrow C_{k}(\mathcal{K}_{j})\).
Then, given two such complexes we can define their \(k\)-th persistent homology group \(H_{k}^{i,j}\) as the image of the homomorphism \(h_{k}^{i,j}\), which is equivalent to the quotient \(\Ker{\partial_{k}^{i}} / (\im{\partial_{k+1}^{j}} \cap \Ker{\partial_{k}^{i}})\) taking \(\Ker{\partial_{k}^{i}}\) as a subgroup of \(\Ker{\partial_{k}^{j}}\).
Persistent homology groups \(H_{k}^{i,j}\) are generated by the \(k\)-dimensional holes in \(\mathcal{K}_{i}\) that are still present in \(\mathcal{K}_{j}\).
The spectral theory in \cite{wang2020persistent, memoli2022persistent} introduces a persistent combinatorial Laplacian or Hodge Laplacian \(\mathcal{L}_{k}^{i,j} = {\partial_{k}^{i}}^{*} \partial_{k}^{i} + \tilde{\partial}_{k+1}^{i,j} { \tilde{\partial}_{k+1}^{i,j} }^{*}\), where \(\tilde{\partial}_{k+1}^{i,j}\) is the restriction of the boundary map \(\partial_{k+1}^{j}\) to the subgroup \(\tilde{C}_{k+1}^{i,j} = \left\{ \phi \in C_{k+1}(\mathcal{K}_{j}) : \partial_{k+1}^{j} \phi \in C_{k}(\mathcal{K}_{i}) \right\}\).
Eigenvalues and eigenspaces of \(\mathcal{L}_{k}^{i,j}\) hold information about the topology and geometry of the corresponding simplicial complexes \(\mathcal{K}_{i}\) and \(\mathcal{K}_{j}\), in particular the dimension of its kernel is the same as the dimension of the persistent homology group \(H_{k}^{i,j}\).

\subsection{Persistence Diagrams}

\begin{figure}[ht]
\centering

  \begin{subfigure}[b]{0.475\textwidth}
  \centering
    \includegraphics[width=0.7\textwidth]{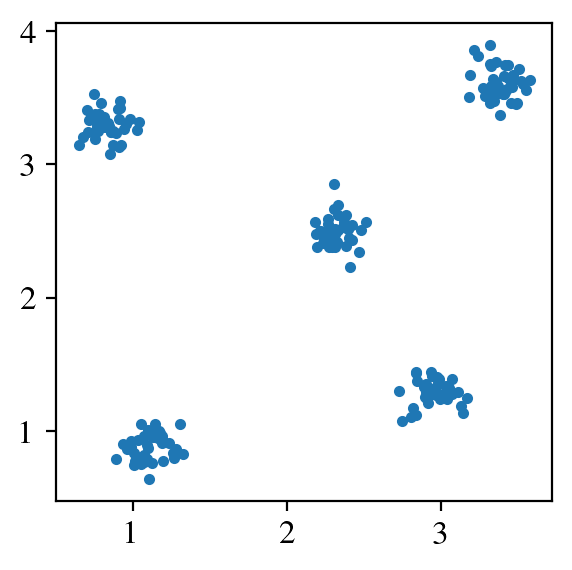}
    \caption{}
    \label{fig:clusters-data}
  \end{subfigure}
  \begin{subfigure}[b]{0.475\textwidth}
  \centering
    \includegraphics[width=0.7\textwidth]{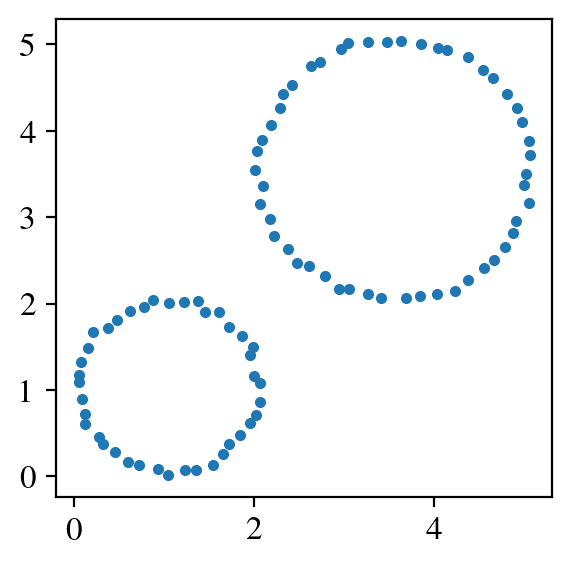}
    \caption{}
    \label{fig:circles-data}
  \end{subfigure}

  \medskip
  \begin{subfigure}[b]{0.475\textwidth}
  \centering
    \includegraphics[width=0.8\textwidth]{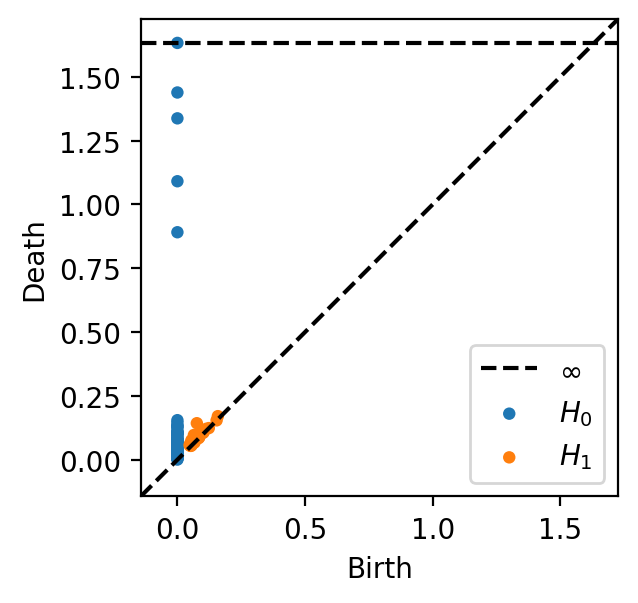}
    \caption{}
    \label{fig:clusters-diagram}
  \end{subfigure}
  \begin{subfigure}[b]{0.475\textwidth}
  \centering
    \includegraphics[width=0.8\textwidth]{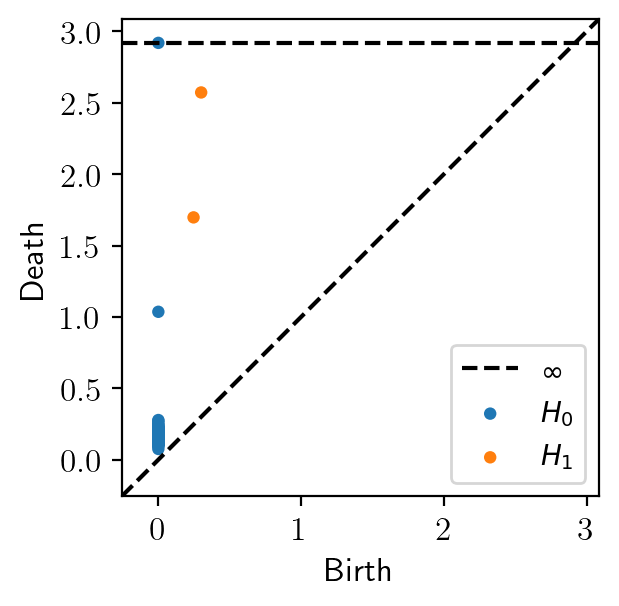}
    \caption{}
    \label{fig:circles-diagram}
  \end{subfigure}

  \caption{
    Examples of persistence diagrams for point cloud data sets.
  }
  \label{fig:diagrams}

\end{figure}

Persistence information can be used to construct persistence diagrams that consist of points \((b_{\eta}, d_\eta)\), where \(b_\eta\) is the scale at which feature \(\eta\) first appears in the filtration, and \(d_\eta\) is the last scale at which it is present.
See Figures \ref{fig:clusters-diagram} and \ref{fig:circles-diagram} for examples.

Persistence diagrams can be useful for classification of their corresponding data sets since they summarize their shape while being invariant to any continuous transformations such as rotating, stretching or shrinking of the data.
For example, the data in Fig. \ref{fig:clusters-data} clusters around five different locations and this information can be recovered from the five connected components (\(H_{0}\)) away from the diagonal in the corresponding persistence diagram in Fig. \ref{fig:clusters-diagram}.
On the other hand, Fig. \ref{fig:circles-data} shows a data set arranged into two separate circles while the corresponding persistence diagram in Fig. \ref{fig:circles-diagram} has two connected components (\(H_{0}\)) and two one-dimensional holes (\(H_{1}\)) away from the diagonal.
Of course, in order to do this one must define a distance on the space of persistence diagrams.
The most popular such distance is the Wasserstein distance \cite{CompyTopo, carlsson2021topological} defined below.

\begin{defn}\label{def:was}
  Let \(\mathcal{D}_{1}, \mathcal{D}_{2}\) be two persistence diagrams, and \(\Delta_{1}, \Delta_{2}\) be their projections to the diagonal.
  Then, their Wasserstein distance is defined as

  \begin{equation}
    d_{p}^W (\mathcal{D}_{1}, \mathcal{D}_{2}) = \inf_{\phi :\tilde{\mathcal{D}_{1}} \leftrightarrow \tilde{\mathcal{D}_{2}}} \left( \sum_{x \in \tilde{\mathcal{D}_{1}}} \|x - \phi(x) \|_{q}^{p} \right)^{\frac{1}{p}}
  \end{equation}

  \noindent where \(q\) is usually chosen as \(\infty\), i.e. the supremum norm, and \(\phi\) are bijections between \(\tilde{\mathcal{D}_{1}} = \mathcal{D}_{1} \cup \Delta_{2}\) and \(\tilde{\mathcal{D}_{2}} = \mathcal{D}_{2} \cup \Delta_{1}\).
\end{defn}

\begin{figure}[ht]
\centering

  \begin{subfigure}[b]{0.475\textwidth}
  \centering
    \includegraphics[width=0.9\textwidth]{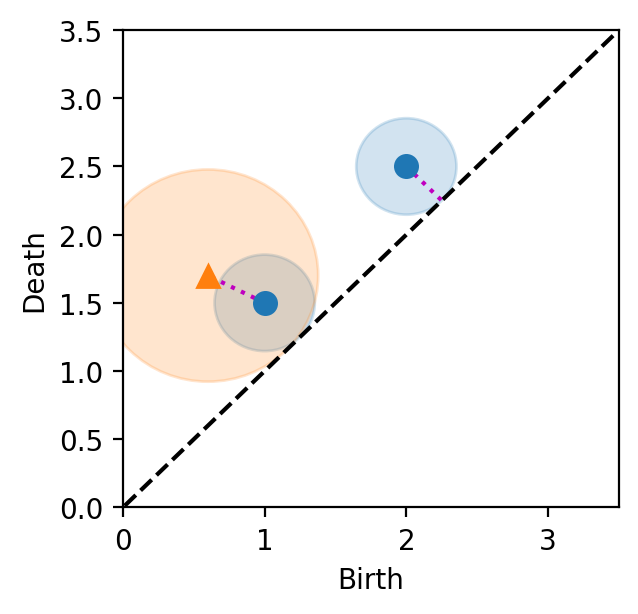}
    \caption{}
    \label{fig:wass-distance}
  \end{subfigure}
  \begin{subfigure}[b]{0.475\textwidth}
  \centering
    \includegraphics[width=0.9\textwidth]{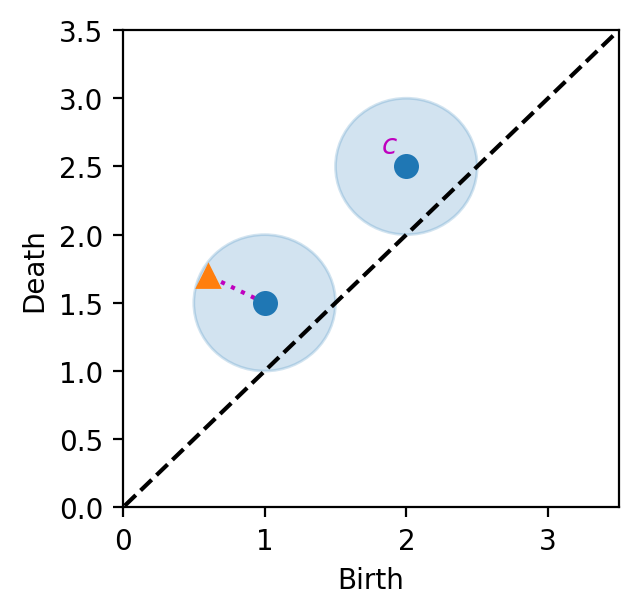}
    \caption{}
    \label{fig:dcp-distance}
  \end{subfigure}

  \caption{
    Example of the Wasserstein (a) and \(d^{c}_{p}\) (b) distances between a persistence diagram with two points (blue dots) and another with one point (orange triangle).
    Disks around the points illustrate the penalization mechanism of each distance.
    The optimal matching in each case is given by the dashed lines.
  }
  \label{fig:distances}

\end{figure}

Based on Definition \ref{def:was}, the Wasserstein distance computes the optimal  matching of the points in two diagrams by penalizing any unmatched points with their distance from the diagonal.
Figure \ref{fig:wass-distance} shows two persistence diagrams, one consists of two blue dots and the other of a single orange triangle.
In addition, each point has a corresponding disk with a radius equal to its distance from the diagonal, which represents the penalty for unmatched points. 
Since two disjoint disks are equivalent to the distance between the corresponding points being larger than their combined distances to the diagonal, one can disregard any bijections \(\phi\) in Def. \ref{def:was} that match points with disjoint disks.
Notice that one could have two different diagrams that consist of completely disjoint disks so that all of their points are matched to the diagonal, however if all of these points are also very close to the diagonal then the Wasserstein distance would be rather small even though the diagrams are different.
This could play an important role in supervised or unsupervised learning \cite{kmeans}, so one could instead consider the following distance in such a case as it was introduced in \cite{Marchese2018}.

\begin{defn}\label{def:dcp}
Consider two persistence diagrams \(\mathcal{D}_{1}, \mathcal{D}_{2}\) with respective cardinalities \(n, m\) such that \(n \le m\).
  Then, their \(d_{p}^{c}\) distance is defined as

  \begin{equation}
    d_{p}^{c}(\mathcal{D}_{1}, \mathcal{D}_{2}) = \left( \frac{1}{m} \left( \inf_{\phi :\mathcal{D}_{1} \hookrightarrow \mathcal{D}_{2}}  \sum_{x \in \mathcal{D}_{1}} \min \left( c, \|x - \phi(x) \|_{q}\right)^{p} + c^{p}|n - m| \right) \right)^{\frac{1}{p}}
  \end{equation}

  \noindent where \(q\) is usually chosen as \(\infty\), i.e. the supremum norm, and \(\phi\) are one-to-one functions from \(\mathcal{D}_{1}\) to \(\mathcal{D}_{2}\).
\end{defn}

The \(d^{c}_{p}\) distance still seeks to match points from two persistence diagrams such that the distance is minimized, but it penalizes unmatched points with a constant value \(c\) rather than their distance to the diagonal.
Fig.~\ref{fig:dcp-distance} shows the same two persistence diagrams that appear in Figure \ref{fig:wass-distance}, but now there are only two disks of radius \(c\) around the blue points.
In order to find the optimal matching that yields the \(d^{c}_{p}\) distance, it suffices to consider only the functions \(\phi\) in Def.~\ref{def:dcp} that match points from the largest diagram to points that lie within their corresponding disk of radius \(c\).
In addition, any points that are isolated with respect to the penalizing constant \(c\) can be completely ignored when computing the distance and simply added later.
Notice that choosing a large enough \(c\) solves the issue that the Wasserstein distance has when points in the diagrams are close to the diagonal.
Moreover, the results provided in \cite{Marchese2018} show that its accuracy for classification tasks is comparable or better (in the data problems of that manuscript) than that of the Wasserstein distance.

\subsection{Quantum Computing Basics}

Quantum theory relies on linear algebra of complex vector spaces.
In particular, a quantum system is described by a Hilbert space \(\HH\) over the complex numbers.
Vectors of the Hilbert space \(\HH\) are called quantum states and are written using a ket \(\ket{\cdot}\) as \(\ket{\psi} \in \HH\).
We may write any quantum state \(\ket{\psi}\) as a linear combination with respect to the standard basis \(\{\ket{\bm{k}}\}_{k}\)
\begin{equation}
  \label{eq:superposition}
  \ket{\psi} = \sum_{i \in k} \psi_{i} \ket{\bm{i}},
\end{equation}
where the coefficients \(\psi_{i} \in \C\) called amplitudes are normalized so that \(\sum_{i} |\psi_{i}|^{2} = 1\).
Linear combinations of quantum states like Eq. \eqref{eq:superposition} with at least two \(\psi_i \neq 0\) are called quantum superpositions.
Elements of the dual space are represented using bra \(\bra{\cdot}\) notation, for instance the dual vector induced by \(\ket{\psi}\) from Eq. \eqref{eq:superposition} is \(\bra{\psi} = \psi_{0}^{*} \bra{\bm{0}} + \psi_{1}^{*} \bra{\bm{1}} + \dots\), with \(\psi_{k}^{*}\) denoting the complex conjugate.

Classical computers have a fundamental unit of information called bit, which can take a value of 0 or 1.
Similarly, quantum computers consider a fundamental unit of information called quantum bit or qubit.
A qubit is the simplest example of a quantum system and is described by the Hilbert space over the complex numbers with standard basis \(\{\ket{0}, \ket{1}\}\).
While a qubit is generally in a state given by a vector \(\ket{\psi} = \alpha \ket{0} + \beta \ket{1}\), an observer cannot see this superposition and instead must perform a measurement to extract useful information.
Measuring \(\ket{\psi}\) with respect to the standard basis for example will result in an outcome of 0 with probability \(|\alpha |^{2}\), or an outcome of 1 with probability \(|\beta |^{2}\).
However, the state of a qubit collapses after a measurement to the state that was observed, so in order to recover a superposition one must prepare it and measure several times.
We can construct larger quantum systems by considering arrays or registers of qubits, which are characterized by the tensor product of the Hilbert spaces for each qubit.
Indeed, a register of \(n\) qubits is described by the Hilbert space over the complex numbers where the standard basis is given by vectors of the form \(\ket{b_{1}} \otimes \ket{b_{2}} \otimes \cdots \otimes \ket{b_{n}}\) with \(b_{k} = 0, 1\).
To avoid cumbersome notation these vectors are often written as \(\ket{b_{1}\dots b_{n}}\) instead, or even just enumerated as \(\ket{\bm{0}}, \dots, \ket{\bm{2^{n} - 1}}\) when the structure of the qubits is not relevant.

The state of a quantum system can be modified via unitary operators.
For example, Pauli operators (\(X, Y, Z\)) are some of the most common single qubit gates, and they can be defined by their actions on the standard basis \(\{\ket{0}, \ket{1}\}\).
The bit-flip operator, \(X\), is the quantum equivalent of the classical \(NOT\) gate, it sends the state \(\ket{0}\) to \(\ket{1}\) and vice versa.
On the other hand, the phase-flip operator, \(Z\), leaves the state \(\ket{0}\) unchanged but flips the sign of the state \(\ket{1}\).
For more details see \cite{quantum-intro-no-physics}.
Quantum algorithms consist of a series of unitary operations \(U_{1}, \dots, U_{n}\) applied to an initial state \(\ket{\psi^{in}}\).
The result is an output quantum state \(\ket{\psi^{out}} = U_{n} U_{n-1} \cdot U_{1} \ket{\psi^{in}}\) which can be either passed to another quantum algorithm or measured.

\subsection{Quantum Persistent Homology}

Since persistent homology is the computational bottleneck of classical TDA methods, a few quantum approaches have appeared in recent years to tackle the problem of extracting persistence information from data sets \cite{ameneyro2024quantum, hayakawa2022quantum, mcardle2022streamlined}.
These algorithms aim to estimate the dimension of the kernel of the persistent combinatorial Laplacian \(\mathcal{L}_{k}^{i,j}\) introduced in \cite{wang2020persistent, memoli2022persistent}.
Our previous work \cite{ameneyro2024quantum} takes a data set \(\{x_{i}\}_{i=1}^{n}\) and considers quantum basis states \(\ket{x_{1}, \dots, x_{n}}\) that encode a simplex \(\sigma = [v_{0}, \dots, v_{k}]\) by setting all qubits \(x_{i}\) that correspond to a vertex \(v_{j}\) in the simplex to state \(\ket{1}\) and all others to state \(\ket{0}\).
This requires only \(\mathcal{O}(n)\) qubits to encode all \(2^{n}\) possible simplices, in contrast to classical algorithms that would need \(\mathcal{O}(2^{n})\) bits instead.
Moreover, the Hilbert space structure of the quantum system \(\mathcal{H}\) provides a natural encoding for the chain groups as vector spaces over \(\mathbb{C}\).
Indeed, we show there are closed subspaces of \(\mathcal{H}\) that encode the chain groups \(C_{k}^{i}\), \(C_{k-1}^{i}\) and subgroups \(\tilde{C}_{k+1}^{i,j}\) on which the persistent combinatorial Laplacians \(\mathcal{L}_{k}^{i,j}\) are defined.
Then we implement the projections onto the desired closed subspaces to restrict the fermionic representation of the boundary operator introduced in \cite{ubaru2021quantum, fermionicBoundary}.
This quantum boundary map \(\partial\) and its adjoint \(\partial^{*}\) are defined on \(\mathcal{H}\) as linear combinations of direct products of Pauli gates.
We define a persistent Dirac operator which can be visualized as the block matrix
\begin{equation}
\label{eq:persistent-dirac}
  B_{k}^{i, j} = \begin{pmatrix}
      0 & P_{k-1}^{i} \partial P_{k}^{i} & 0 \\
      P_{k}^{i} \partial^{*} P_{k-1}^{i} & 0 & P_{k}^{i} \partial P_{k+1}^{j} \\
      0 & P_{k+1}^{j} \partial^{*} P_{k}^{i} & 0
  \end{pmatrix},
\end{equation}
and we prove that the positive spectra of a shifted version \(B_{k}^{i,j}[\xi]\) of this persistent Dirac operator corresponds to the spectrum of the persistent combinatorial Laplacian \(\mathcal{L}_{k}^{i,j}\).
Therefore, one may use quantum phase estimation or other similar quantum techniques to recover the positive spectra of the shifted Dirac operator \(B_{k}^{i,j}[\xi]\) and use it to deduce the persistence information of a data set along with other non-harmonic features of the persistent Laplacian.
While this method could provide a quadratic or Grover-like speedup over equivalent classical algorithms and yield more information than other quantum algorithms for persistent homology \cite{hayakawa2022quantum, mcardle2022streamlined}, the techniques utilized to implement the projections and estimate the spectra of the persistent Dirac operator require fault tolerant quantum computers.

\subsection{QAOA}

Since current and near future quantum hardware is not fault tolerant, classic-quantum hybrid algorithms are often more attractive than pure quantum ones.
The Quantum Approximate Optimization Algorithm (QAOA) \cite{qaoa} is such an example of a hybrid algorithm.
QAOA approximates the ground state of a Hamiltonian by alternating two unitary operators \(U_{C}\) and \(U_{M}\) applied to a suitable initial state.
The unitary \(U_{C} = e^{-i\gamma H_{C}}\) is defined by the problem Hamiltonian \(H_{C}\) and a parameter \(\gamma \in [0, 2\pi)\), while the other operator \(U_{M} = e^{-i\beta H_{M}}\) consists of a mixing Hamiltonian \(H_{M}\) and a parameter \(\beta \in [0, 2\pi)\).

Here, the problem Hamiltonian \(H_{C}\) encodes the cost of the optimization problem to be solved, while the mixing Hamiltonian \(H_{M}\) is used to produce a superposition over all the quantum states of interest (those encoding possible solutions).
After \(p\) iterations of the unitaries applied to the initial state, one obtains the trial state \(|\psi (\beta, \gamma)\ra_p\).
The goal of the algorithm is to find parameters \(\beta\) and \(\gamma\) that maximize the expected value of the problem Hamiltonian: \(_{p}\la \psi (\beta, \gamma)| H_{C} |\psi (\beta, \gamma)\ra_p\).
QAOA has been used to solve combinatorial optimization problems over binary variables, such as MaxCut problems which involve partitioning the vertices of a graph into two sets in a way that maximizes the number of edges between sets \cite{qaoaWeights, qaoaGraphMatching}.
Section \ref{sec:methods} displays the process of how computing the distance between two persistence diagrams can be viewed as a combinatorial optimization problem over binary variables.

\section{Methods} \label{sec:methods}

In this section, we formulate combinatorial optimization problems over binary variables to compute the distances defined earlier.
Then we construct quantum algorithms to estimate the solution to the optimization problems.

\subsection{Quantum Algorithm for Wasserstein Distance}
\label{sec:was-optimization-problem}

\begin{figure}
\centering
  \begin{tikzpicture}[
    point1/.style={circle, draw=green!60, fill=green!5, very thick, minimum size=7mm},
    aux1/.style={rectangle, draw=red!60, fill=red!5, very thick, minimum size=5mm},
    point2/.style={circle, draw=blue!60, fill=blue!5, very thick, minimum size=7mm},
    aux2/.style={rectangle, draw=orange!60, fill=orange!5, very thick, minimum size=5mm},
    ]
    \node[point1] (x1)                {\(x_{1}\)};
    \node[point2] (y1)  [right=2cm of x1] {\(y_{1}\)};
    \node[point1] (dx1) [below=of x1] {\(\cdots\)};
    \node[point2] (dy1) [below=of y1] {\(\cdots\)};
    \node[point1] (xn)  [below=of dx1] {\(x_{n}\)};
    \node[point2] (ym)  [below=of dy1] {\(y_{m}\)};
    \node[aux1]   (xa1) [left=2cm of x1] {\(\tilde{x_{1}}\)};
    \node[aux2]   (ya1) [right=2cm of y1] {\(\tilde{y_{1}}\)};
    \node[aux1]   (dx2) [left=of dx1] {\(\cdots\)};
    \node[aux2]   (dy2) [right=of dy1] {\(\cdots\)};
    \node[aux1]   (xan) [left=of xn] {\(\tilde{x_{n}}\)};
    \node[aux2]   (yam) [right=of ym] {\(\tilde{y_{m}}\)};

    \draw[-] (x1.west) -- node[above] {\tiny \(\|Px_{1} - x_{1}\|_{q}^{p}\)} (xa1.east);
    \draw[-] (x1.east) -- node[above] {\tiny \(\|x_{1} - y_{1}\|_{q}^{p}\)} (y1.west) ;
    \draw[-] (y1.east) -- node[above] {\tiny \(\|y_{1} - Py_{1}\|_{q}^{p}\)} (ya1.west) ;
    \draw[-] (x1.east) -- (dy1.west);
    \draw[-] (x1.east) -- (ym.west) ;
    \draw[-] (xn.east) -- (y1.west) ;
    \draw[-] (xn.east) -- (dy1.west);
    \draw[-] (xn.east) -- (ym.west) ;
    \draw[-] (xn.west) -- (xan.east);
    \draw[-] (dx1.east) -- (y1.west) ;
    \draw[-] (dx1.east) -- (dy1.west);
    \draw[-] (dx1.east) -- (ym.west) ;
    \draw[-] (dx2.east) -- (dx1.west);
    \draw[-] (dy2.west) -- (dy1.east);
    \draw[-] (yam.west) -- (ym.east) ;
  \end{tikzpicture}
  \caption{Graph connecting the points of two persistence diagrams.} 
  \label{fig:was-graph}
\end{figure}
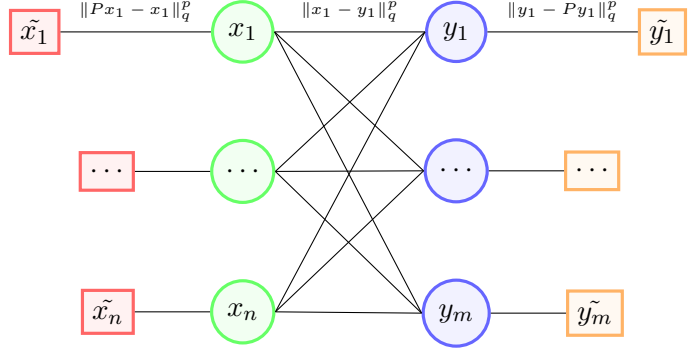

We begin by creating an optimization problem that encodes the Wasserstein distance between persistence diagrams introduced in Def. \ref{def:was}.
Let \(\mathcal{D}_{1}, \mathcal{D}_{2}\) be persistence diagrams and consider a weighted graph \(G = (V, E)\).
The vertex set \(V\) consists of the points in \(x_{i} \in \mathcal{D}_{1}\) and \(y_{j} \in \mathcal{D}_{2}\) along with an auxiliary vertex \(\tilde{x_{i}}\) or \(\tilde{y_{j}}\) for each point.
These auxiliary vertices represent the projection of a point to the diagonal of the diagram and are meant to penalize unmatched points when the diagrams have different cardinality.
The edge set \(E\) on the other hand is formed by all the possible edges \((x_{i}, y_{j})\) between points \(x_{i} \in \mathcal{D}_{1}\) and \(y_{j} \in \mathcal{D}_{2}\), as well as the edges between each point and their auxiliary vertex \((x_{i}, \tilde{x}_{i})\) or \((\tilde{y}_{j}, y_{j})\). A \textit{main edge} is defined to be an edge connecting some \(x_{i} \in \mathcal{D}_{1}\) and \(y_{j} \in \mathcal{D}_{2}\).
The main edges produce a complete bipartite graph between the points in the persistence diagrams with leaves to vertices that represent the closest point on the diagonal to each point.
See Fig. \ref{fig:was-graph} for a visual representation of this graph.

Consider the optimization problem over binary variables

\begin{equation}\label{eq:cost-was}
  \text{minimize } \ C = \sum_{e \in E} w_{e} \delta_{e}
\end{equation}
\begin{equation}\label{eq:constraints-was}
  \text{subject to } \ \sum_{e \sim v} \delta_{e} = 1 \ , \ \ \forall v \in \mathcal{D}_{1}, \mathcal{D}_{2} 
\end{equation}
where \(\delta_{e} \in \{0, 1\}\) is \(1\) if an edge is in a matching and \(0\) if not, \(w_{e}\) are the weights of the edges, and \(e \sim v\) denotes the edges incident to vertex \(v\).

\begin{lemma}\label{lem:waserstein}
  The solution to the optimization problem as defined in Eqs. \eqref{eq:cost-was} and \eqref{eq:constraints-was} using the weights
  \begin{equation}\label{eq:we-was}
    w_{e} =
    \begin{cases}
      \|x_{i} - y_{j}\|_{q}^{p} & e = (x_i, y_j) \\
      \|x_{i} - Px_{i}\|_{q}^{p} & e = (x_i, \tilde{x}_i) \\
      \|Py_{j} - y_{j}\|_{q}^{p} & e = (\tilde{y}_j, y_j)
    \end{cases}
  \end{equation}
  is the distance \(d_p^W (\mathcal{D}_1, \mathcal{D}_2)\) introduced in Def. \ref{def:was}.
  Here, \(P x = \frac{1}{2} (a + b, a + b)^{T}\) denotes the projection to the diagonal of \(x = (a, b)^{T}\).

  \begin{proof}
    The cost function in Eq. \eqref{eq:cost-was} represents the sum in Def. \ref{def:was}, while the constraints in Eq. \eqref{eq:constraints-was} ensure that each vertex is matched exactly once in order to encode the bijections between \(\tilde{\mathcal{D}_{1}}\) and \(\tilde{\mathcal{D}_{2}}\).
    Notice that the auxiliary vertices represent the projection of each point to the diagonal, and the weight that is assigned to the corresponding edges is simply the distance to the diagonal.
    For a choice of \(\delta_e\)'s that satisfies the constraints and with \(\phi\) the bijection that corresponds to it, the cost in Eq. \eqref{eq:cost-was} is simply \(\text{Cost} [\phi] = \sum_{x\in \tilde{\mathcal{D}}_1} \|x - \phi(x)\|_q^p\).
    Since taking the minimum over the different solutions to the constraints, Eq. \eqref{eq:constraints-was} is the same as taking the minimum over the bijections, and thus the result of the optimization problem must be \(d_p^W\). 
  \end{proof}
\end{lemma}


Now we describe the quantum algorithm to compute the optimization problem within a quantum framework given in Eq. \eqref{eq:cost-was} with the constraints as in Eq. \eqref{eq:constraints-was}.
We use qubits \(|e\ra\) to encode the edges \(e\in E\) such that the state of \(|e\ra\) corresponds to the binary variable \(\overline{\delta_{e}} = 1 + \delta_{e} \mod 2\).
Thus a qubit \(|e\ra\) in state \(|0\ra\) means that edge \(e\) is present.
In general we can use \(n \times m \) qubits to identify the edges between points in the diagrams, along with \(n + m\) more qubits for the edges to auxiliary vertices.
The quantum states \(\bigotimes_{e \in E}|e\ra\) encode all the different sub-graphs of \(G\).
Throughout this paper, the quantum states are  arranged as in Eq. \eqref{eq:qstates-was}, where each row will correspond to a point in \(\mathcal{D}_{1}\) and each column to a point in \(\mathcal{D}_{2}\).
In particular this makes it easier to recover the respective matching.

\begin{equation}\label{eq:qstates-was}
  \ket*{
    \begin{array}{cccc}
      (x_{1}, y_{1}) & \cdots & (x_{1}, y_{m}) & (x_{1}, \tilde{x}_{1}) \\
      \vdots & \ddots & \vdots & \vdots \\
      (x_{n}, y_{1}) & \cdots & (x_{n}, y_{m}) & (x_{n}, \tilde{x}_{n}) \\
      (\tilde{y}_{1}, y_{1}) & \cdots & (\tilde{y}_{m}, y_{m}) &
    \end{array}
  }
\end{equation}

In order to construct a Hamiltonian that encodes the cost function in Eq. \eqref{eq:cost-was} we need to modify it to use the variables \(\overline{\delta}_{e}\) that correspond to the states of the qubits.
Notice that \( C = \sum_{e\in E}w_{e} - \sum_{e\in E} w_e \overline{\delta_{e}}\), where \(w_{e}\) are the weights introduced in Eq. \eqref{eq:we-was}.
Since the first term is constant, minimizing \(\text{Cost}[\phi]\) is equivalent to maximizing \(g = \sum_{e\in E} w_{e} \overline{\delta_{e}}\).
So we may use the cost Hamiltonian \( H_{g} =  \frac{1}{2} \sum_{e \in E} w_{e} (I - Z_e) \) whose ground state maximizes \(g\).
The introduction of weights into the cost Hamiltonian \(H_{g}\) inspired by \cite{qaoaWeights} is essential to capture the distances between points of the persistence diagrams.
This introduction of weights can be recast as other QAOA variants, such as multi-angle QAOA \cite{herrman2022multi}.
This yields our unitary cost operator

\begin{equation}\label{eq:uni-cH}
  U_{g} = e^{i \frac{\gamma}{2} \sum_{e \in E} w_{e} Z_e} = \prod_{e \in E} e^{i \frac{\gamma}{2} w_{e} Z_e} =  \prod_{e \in E} R_{Z_e} (-\gamma w_{e})\ ,
\end{equation}
where \(Z_{e}\) and \(R_{Z_e} (-\gamma w_e)\) are the Pauli-\(Z\) operator and the single qubit rotation over the \(z\)-axis, respectively, acting on qubit \(|e\ra\).
We omit the constant term \(\frac{1}{2} \sum_{e\in E}w_{e} I \) in the Hamiltonian since it only adds a global phase \(e^{-i \frac{1}{2}\sum_{e\in E}w_{e}}\) to the unitary operator.

To encode the constraints in Eq. \eqref{eq:constraints-was}, we use control clauses \(f\) as in \cite{qaoaGraphMatching}, to build an individual mixing Hamiltonian \(M_{e} = f(e) X_{e}\) for each edge, where \(X_{e}\) is the Pauli-\(X\) operator acting on qubit \(|e\ra\).

\begin{defn}
\label{def:control-clause-wass}
We define the control clause \(f\) as the binary function on the set of edges \(E\) such that
  \begin{equation}\label{eq:control-was}
    f(e) =
    \begin{cases}
      \overline{\delta_{e}} \prod_{l \ne i} \overline{\delta_{x_l, y_j}} \prod_{k \ne j} \overline{\delta_{x_i, y_k}} & e = (x_{i}, y_{j}) \\
      \delta_{e} \overline{\prod_{k = 1}^{|\mathcal{D}_{2}|} \overline{\delta_{x_i, y_k}}} & e = (x_{i}, \tilde{x}_{i}) \\
      \delta_{e} \overline{\prod_{l = 1}^{|\mathcal{D}_{1}|} \overline{\delta_{x_l, y_j}}} & e = (\tilde{y}_{j}, y_{j})
    \end{cases}.
  \end{equation}
\end{defn}

Notice that for an edge \(e = (x_{i}, y_{j})\), the control clause \(f(e)\) will be 1 whenever both points \(x_{i}\) and \(y_{j}\) are unmatched.
On the other hand, for an edge to an auxiliary vertex \(e = (x_{i}, \tilde{x}_{i})\) the control clause \(f(e)\) will be 0 either when the point \(x_{i}\) is unmatched, or when its corresponding auxiliary vertex \(\tilde{x}_{i}\) is unmatched.
Similarly, for an edge \(e = (\tilde{y}_{j}, y_{j})\), \(f(e)\) is 0 if either \(y_{j}\) or \(\tilde{y}_{j}\) are unmatched.
So this control clause captures the idea behind our constraints of mathcing each point exactly once.

We use the individual mixing Hamiltonians \(M_{e}\) to build an individual mixing unitary operator for each edge
\begin{equation}\label{eq:mixing-e-was}
  U_{M, e}(\beta)  = e^{-i\beta M_{e}}  = \bigwedge_{f(e)} e^{-i\beta X_{e}}  = \bigwedge_{f(e)} R_{X_{e}}(\beta),
\end{equation}
and these individual unitaries are applied in succession to build the complete mixing unitary operator
\begin{equation}\label{eq:mixing-unitary-was}
  U_{M}(\beta) = \prod_{e \in E} U_{M, e}(\beta) = \prod_{e\in E} e^{-i\beta M_{e}} = \prod_{e\in E} \bigwedge_{f(e)} R_{X_{e}}(\beta) .
\end{equation}

Note however that the order in which one applies the individual operators matters, we apply the operators corresponding to the main vertices first and those corresponding to auxiliary vertices last.
Our goal is to create a mixing operator that produces only quantum states that satisfy the constraints of the problem. 
But the mixing unitary operator in Eq. \eqref{eq:mixing-unitary-was} creates some quantum states that don't satisfy the constraints in Eq. \eqref{eq:constraints-was}, i.e. matchings that don't represent bijections.
So, we must show that all of these problematic quantum states don't affect the optimization process.
Indeed, as stated in Thm. \ref{thm:feasibility-was}, we may relax the constraints of the problem so that any extra quantum states produced by our mixing unitary satisfy these new constraints.
Moreover, matchings corresponding to extra quantum states are the result of adding edges from auxiliary vertices to the matchings that do represent bijections.
In particular, their cost will be greater and thus the minimum will remain the same.

\begin{theorem}\label{thm:feasibility-was}
  The mixing operator \(U_M (\beta)\) preserves the feasibility of quantum states.
  That is, if the quantum state \(\ket{s}\) represents a matching that satisfies the constraints
  \begin{equation}\label{eq:relaxed-constraints-was}
    \begin{split}
      \sum_{(i,j) \sim v} \delta_{(i,j)} \le 1 \ &, \ \ \mbox{for all } v \in \mathcal{D}_{1}, \mathcal{D}_{2} \\
      \sum_{e \sim v} \delta_{e} \ge 1 \ &, \ \ \mbox{ for all } v \in \mathcal{D}_{1}, \mathcal{D}_{2}
    \end{split}
  \end{equation}
  then \(U_M (\beta) \ket{s}\)  also satisfies these constraints.
  Moreover, the minimum of the cost function in Eq. \eqref{eq:cost-was} over these constraints is the same as the minimum over the constraints in Eq. \eqref{eq:constraints-was}.

  \begin{proof}
    We first give some intuition about these new constraints and show that the minimum remains the same.
    Notice that the first constraint asks that a point \(v\) in one of the persistence diagrams is matched at most once to points in the other diagram.
    The second constraint requires that every point in the persistence diagrams is matched at least once.
    This means that points in a persistence diagram \(\mathcal{D}_{i}, i \in \{1,2\}\) could be incident to more than one edge, but only one of the edges will connect it to a point in the other diagram and all other edges connect to auxiliary vertices.
    So the solutions to these constraints will be matchings that represent a bijection or matchings that result of adding connections to auxiliary vertices from a bijection.
    In particular, any solution to the constraints in Eq. \eqref{eq:constraints-was} is also a solution to Eq. \eqref{eq:relaxed-constraints-was}, and any solution to the latter that is not also a solution to the former must have a larger cost.

    Now, given a quantum state \(\ket{s}\) which satisfies the constraints in Eq. \eqref{eq:relaxed-constraints-was}, we show that \(U_{M,e}(\beta) \ket{s}\) also satisfies the constraints for arbitrary \(e\in E\).
    This in turn proves that \(U_M(\beta)\) preserves the feasibility of quantum states as it is defined as a product of the individual unitaries.

    From Eq. \eqref{eq:mixing-e-was} note that there are only two possible cases for \(U_{M,e}(\beta) \ket{s}\) depending on the value of \(f(e)\).
    First, if \(f(e) = 0\) the output state is simply \(\ket{s}\).
    On the other hand, when \(f(e) = 1\) the resulting state is \(\cos\beta I \ket{s} - i \sin\beta X_{e} \ket{s}\).
    The first term is just \(\ket{s}\) scaled by a constant, so we only need to verify that the second term is feasible.

    Notice that the second term \(X_{e}\) flips the qubit \(|e\ra\) but it is only applied when \(f(e) = 1\).
    If \(e = (x_{i},y_{j})\) is an edge between points in the diagrams, \(f(e) = 1\) if and only if \(\delta_{e}, \delta_{i,k}, \delta_{l,j} = 0\) for all \(l \ne i\) and \(k \ne j\), that is, the edge \(e = (x_{i}, y_{j})\) is added to the matching if there are no other main edges containing vertices \(x_{i}\) or \(y_{j}\).
    Since this operation adds an edge, the second constraint in Eq. \eqref{eq:relaxed-constraints-was} is trivially satisfied, moreover the first constraint is satisfied because the edge is only added if that sum is equal to zero.
    On the other hand, if \(e = (x_{i}, \tilde{x}_{i})\) (or \((\tilde{y}_{j}, y_{j})\)) is an edge to an auxiliary vertex, \(f(e) = 1\) whenever \(\delta_{e} = 1\) and at least one of \(\delta_{i,k}\) for \(1 \le k \le |\mathcal{D}_{2}|\) (or \(\delta_{l,j}\) for \(1 \le l \le |\mathcal{D}_{1}|\)) is non-zero, in other words, we remove the edge \(e = (x_{i}, \tilde{x}_{i})\) (or \((\tilde{y}_{j}, y_{j})\)) from the matching if there is a main edge connecting the corresponding vertex \(x_{i}\) (or \(y_{j}\)) to a vertex in the other diagram.
    So, the first constraint in Eq. \eqref{eq:relaxed-constraints-was} is unaffected and the second one is still satisfied as we only remove an edge if the corresponding sum is at least 2.
    All in all, \(U_{M,e}(\beta) \ket{s}\) is a superposition of quantum states whose corresponding matchings satisfy the constraints in Eq. \eqref{eq:relaxed-constraints-was}.
  \end{proof}
\end{theorem}

Another important property of our mixing operator is that it produces every solution to the relaxed constraits in Eq. \eqref{eq:relaxed-constraints-was} in a single iteration.
Using the mixing operator again will change the amplitudes of the quantum states but it won't create any new states.
For this reason we also use the mixing operator to initialize the algorithm with a superposition of only the necessary quantum states to solve the optimization problem.

\begin{theorem}\label{thm:completeness-was}
  Applying each of the \(n \times m + n + m\) unitaries \(U_{M,e} (\beta)\) in sequence starting with the initial state
  \begin{equation}\label{eq:trivial-was}
    \ket*{
      \begin{array}{cccc}
        1 & \cdots & 1 & 0 \\
        \vdots & \ddots & \vdots & \vdots \\
        1 & \cdots & 1 & 0 \\
        0 & \cdots & 0 &
      \end{array}
    }
  \end{equation}
  yields a superposition of all possible matchings that satisfy Eq. \eqref{eq:relaxed-constraints-was} with non-zero amplitudes.
  \begin{proof} 
    Notice that one may obtain all solutions to Eq. \eqref{eq:relaxed-constraints-was} by finding every possible choice for the main \(n \times m\) edges that satisfy the first constraint, and then for each of these obtaining the different combinations of the \(n + m\) auxiliary edges that satisfy the second constraint.
    So, we first apply the unitaries corresponding to the main edges to produce the solutions to the first constraint while keeping all the auxiliary edges on.
    After which we use the remaining unitaries to get the combinations of auxiliary edges.
    Since every unitary keeps the quantum state it acts on, either whole or scaled by \(\cos\beta\), this process yields all matchings that satisfy Eq. \eqref{eq:relaxed-constraints-was}.

    Let \(e_1, \dots, e_N\) be any ordering of the \(n\times m\) main edges, we want to prove that applying their corresponding unitaries in sequence will produce a superposition of all solutions to the first constraint in Eq. \eqref{eq:relaxed-constraints-was}.
    We proceed by induction, where we will prove that applying \(U_{M,e_k}\) to the superposition of all feasible matches using only edges \(e_1, \dots, e_{k-1}\), yields a superposition of all feasible matchings using only edges \(e_1, \dots, e_{k}\).
    Since the case for \(k = 1\) starts with the quantum state in Eq. \eqref{eq:trivial-was}, \(f(e_1)\) will always equal 1 and the result of applying \(U_{M,e_1}\) is a superposition of this trivial matching and the matching which includes only edge \(e_1\).
    For \(k > 1\), assume we have a superposition of all feasible matchings with edges \(e_1, \dots, e_{k-1}\) and notice that applying \(U_{M,e_k}\) to any one of these matchings can have two possible outcomes.
    Indeed if \(f(e_k) = 0\) the matching remains unchanged.
    On the other hand, if \(f(e_k) = 1\) the output will be a combination of the input matching and the matching that results from adding edge \(e_k\).
    Therefore, \(U_{M, e_k}\) will retain all the matchings with only edges \(e_1, \dots, e_{k-1}\) and produce all feasible matchings that result from adding edge \(e_k\), which yields a superposition over all feasible matchings that use only edges \(e_1, \dots, e_k\).

    Now let \(e_{1}', \dots, e_{N'}'\) be any ordering of the \(n + m\) auxiliary edges.
    We prove in a similar way that given the superposition that results of the previous step, applying the auxiliary unitaries in sequence yields a superposition of all possible solutions to Eq. \eqref{eq:relaxed-constraints-was}.
    Indeed, applying \(U_{M, e_{1}'}\) to each of the matchings in the superposition from the previous step gives either the same matching when \(f(e_{1}') = 0\) or a combination of the input matching and the one that results of removing edge \(e_{1}'\) when \(f(e_{1}') = 1\), producing all feasible choices for \(e_{1}'\).
    In the same manner, applying \(U_{M, e_{k}'}\) to the superposition with all feasible choices for edges \(e_{1}', \dots, e_{k-1}'\) will produce a superposition with all feasible combinations for edges \(e_{1}', \dots, e_{k}'\).
  \end{proof}
\end{theorem}

\begin{remark}
This quantum algorithm uses \(n \times m + n + m\) logical qubits to encode the edges and  a few extra ancillary qubits for the controls.
One must apply the individual mixing operator for each edge in sequence, which requires \(n \times m + n + m\) multi-qubit controled rotation gates \(R_{X}\), as well as \(n \times m + n + m\) single qubit rotations \(R_{Z}\) for the cost operator.
So the overall cost of the algorithm is \(\mathcal{O}(n m)\) operations, which is better than \(\mathcal{O}(n^{2} m)\) operations needed by the Hungarian algorithm used to compute this distance classically.
\end{remark}


\subsection{A More Efficient Distance}
\label{sec:dcp}

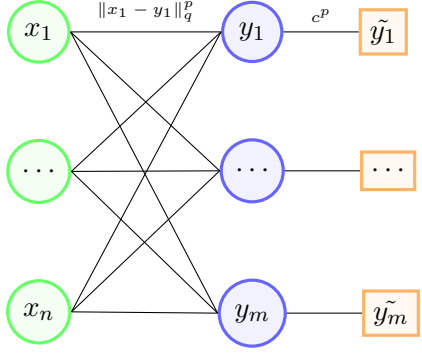
\begin{figure}
\centering
  \begin{tikzpicture}[
    point1/.style={circle, draw=green!60, fill=green!5, very thick, minimum size=7mm},
    aux1/.style={rectangle, draw=red!60, fill=red!5, very thick, minimum size=5mm},
    point2/.style={circle, draw=blue!60, fill=blue!5, very thick, minimum size=7mm},
    aux2/.style={rectangle, draw=orange!60, fill=orange!5, very thick, minimum size=5mm},
    ]
    \node[point1] (x1)                {\(x_{1}\)};
    \node[point2] (y1)  [right=2cm of x1] {\(y_{1}\)};
    \node[point1]   (dx1) [below=of x1] {\(\cdots\)};
    \node[point2]   (dy1) [below=of y1] {\(\cdots\)};
    \node[point1] (xn)  [below=of dx1] {\(x_{n}\)};
    \node[point2] (ym)  [below=of dy1] {\(y_{m}\)};
    \node[aux2]   (ya1) [right=of y1] {\(\tilde{y_{1}}\)};
    \node[aux2]   (dy2) [right=of dy1] {\(\cdots\)};
    \node[aux2]   (yam) [right=of ym] {\(\tilde{y_{m}}\)};

    \draw[-] (x1.east) -- node[above] {\tiny \(\|x_{1} - y_{1}\|_{q}^{p}\)} (y1.west) ;
    \draw[-] (y1.east) -- node[above] {\tiny \(c^{p}\)} (ya1.west) ;
    \draw[-] (x1.east) -- (dy1.west);
    \draw[-] (x1.east) -- (ym.west);
    \draw[-] (xn.east) -- (y1.west);
    \draw[-] (xn.east) -- (dy1.west);
    \draw[-] (xn.east) -- (ym.west);
    \draw[-] (dx1.east) -- (y1.west);
    \draw[-] (dx1.east) -- (dy1.west);
    \draw[-] (dx1.east) -- (ym.west);
    \draw[-] (dy2.west) -- (dy1.east);
    \draw[-] (yam.west) -- (ym.east);
  \end{tikzpicture}
  \caption{Graph connecting the points of two persistence diagrams.}
  \label{fig:dcp-graph}
\end{figure}

We proceed in a similar manner by describing the optimization problem and then building a quantum algorithm to solve it.
Let \(\mathcal{D}_{1}, \mathcal{D}_{2}\) be persistence diagrams such that their cardinalities satisfy the relationship, \(|\mathcal{D}_{1}| \le |\mathcal{D}_{2}|\).
Consider a weighted graph \(G = (V, E)\).
This time the vertex set \(V\) consists of the points in \(x_{i} \in \mathcal{D}_{1}\) and \(y_{j} \in \mathcal{D}_{2}\) along with an auxiliary vertex \(\tilde{y_{j}}\) for each point in \(\mathcal{D}_{2}\).
The auxiliary vertices still serve the purpose of penalizing any unmatched points due to a difference in cardinality, but now they also compute  \(\min (c, \|x_{i} - y_{j}\|_{q})\).
The edge set \(E\) on the other hand is formed by all the possible edges \((x_{i}, y_{j})\) between points \(x_{i} \in \mathcal{D}_{1}\) and \(y_{j} \in \mathcal{D}_{2}\), as well as the edges between each point in \(\mathcal{D}_{2}\) and their auxiliary vertex \((\tilde{y}_{j}, y_{j})\).
This results in a smaller graph, which in turn will reduce the number of qubits and controls needed for the quantum algorithm.
See Fig. \ref{fig:dcp-graph} for a visual representation of this graph.

Now consider the following optimization problem over binary variables.

\begin{equation}\label{eq:cost-dcp}
  \text{minimize } \ C = \sum_{e \in E} w_{e} \delta_{e}
\end{equation}
\begin{equation}\label{eq:constraints-dcp}
  \begin{split}
    \text{subject to } \ \sum_{e \sim v} \delta_{e} &= 1 \ , \ \ \forall v \in \mathcal{D}_{2} \\
    \text{and } \ \sum_{e \sim v} \delta_{e} &\le 1 \ , \ \ \forall v \in \mathcal{D}_{1}
  \end{split}
\end{equation}

The first constraint ensures that every point in \(\mathcal{D}_{2}\) is matched exactly once, either to a point in \(\mathcal{D}_{1}\) or to its auxiliary vertex.
On the other hand, the second constraint allows points in \(\mathcal{D}_{1}\) to be matched once or to remain unmatched. 
These choices represent the different possible outcomes, \(c\) or \(\|x_{i} - y_{j}\|\), of the corresponding minimum that appears in the sum at Def. \ref{def:dcp}.
As before, we show that for a particular choice of weights, the solution to this optimization problem yields the distance \(d^{c}_{p}(\mathcal{D}_{1}, \mathcal{D}_{2})\).

\begin{lemma}\label{lem:dcp}
  The solution to the optimization problem described in Eqs. \eqref{eq:cost-dcp} and \eqref{eq:constraints-dcp} using the weights
  \begin{equation}\label{eq:we-dcp}
    w_{e} =
    \begin{cases}
      \|x_{i} - y_{j}\|_{q}^{p} & e = (x_i, y_j) \\
      c^{p} & e = (\tilde{y}_j, y_j)
    \end{cases}
  \end{equation}
  is the distance \(d_p^c (\mathcal{D}_1, \mathcal{D}_2)\) introduced in Def. \ref{def:dcp}.

  \begin{proof}
    Notice that in order to compute the sum in Def. \ref{def:dcp} one must choose for each point in \(\mathcal{D}_{2}\) wether to match it to an unmatched point in \(\mathcal{D}_{1}\) or penalize it with the parameter \(c\).
    The auxiliary vertices serve the purpose of penalizing any unmatched points in \(\mathcal{D}_2\).
    On the other hand, the constraints ensure that each point in \(\mathcal{D}_{2}\) is matched exactly once, either to a point in \(\mathcal{D}_{1}\) or to an auxiliary vertex, and that each point in \(\mathcal{D}_{1}\) is matched at most once.

    Given a solution to the constraints, an edge of the form \(e = (x_i, y_j)\) with \(\delta_e = 1\) represents matching vertices \(x_i\) and \(y_j\) with weight \(\|x_i - y_j\|_q^p\).
    On the other hand, an edge of the form \(e' = (\tilde{y}_j, y_j)\) represent the vertex \(y_j\) penalized with the weight \(c^p\) instead.

    Then the cost in Eq. \eqref{eq:cost-dcp} will be the sum that appears in Def. \ref{def:dcp} with either \(c\) or \(\|x - \gamma(x)\|_q\) for each point \(x\) instead of the minimum.
    However, since this is a minimization problem, the algorithm will prefer the matching with the weight that corresponds to \(\min (c, \|x - \gamma(x)\|_q)\).
    Thus, minimizing over all solutions to the constraints yields the distance \(d_p^c\).
  \end{proof}
\end{lemma}

We use again qubits \(|e\ra\) to encode the edges \(e\in E\) of the graph.
We still require \(n \times m \) qubits to identify the main edges between points in the diagrams, but now we only need \(m\) more qubits for the edges to auxiliary vertices.
The resulting quantum states \(\bigotimes_{e\in E}|e\ra\) encode all the different sub-graphs of \(G\) and are often arranged throughout this text as in Eq. \eqref{eq:qstates-dcp} below,

\begin{equation}\label{eq:qstates-dcp}
  \ket*{
    \begin{array}{ccc}
      (x_1, y_1) & \cdots & (x_1, y_m)  \\
      \vdots & \ddots & \vdots \\
      (x_n, y_1) & \cdots & (x_n, y_m) \\
      (\tilde{y}_1, y_1) & \cdots & (\tilde{y}_m, y_m)
    \end{array}
  }.
\end{equation}

To construct a Hamiltonian for the cost function in Eq.~ \eqref{eq:cost-dcp}, we must once more transform it into \(g = \sum_{e\in E} w_{e} \overline{\delta_{e}}\) to avoid unnecessary gates.
Note that the weights \(w_{e}\) are now the ones introduced in Eq. \eqref{eq:we-dcp}.
This results in a unitary cost operator of the same form as in Eq. \eqref{eq:uni-cH} for the Wasserstein distance, but with the edges and weights corresponding to the \(d^{c}_{p}\) distance.
To encode the constraints in Eq. \eqref{eq:constraints-dcp} we use similar control clauses to build individual mixing Hamiltonians \(M_{e} = f(e) X_{e}\) for each edge.

\begin{defn}
\label{def:control-clause-dcp}
We define the control clause \(f\) as the binary function on the set of edges \(E\) such that
  \begin{equation}\label{eq:control-dcp}
    f(e) =
    \begin{cases}
      \overline{\delta_{e}} \prod_{l \ne i} \overline{\delta_{x_l, y_j}} \prod_{k \ne j} \overline{\delta_{x_i, y_k}} & e = (x_{i}, y_{j}) \\
      \delta_{e} \overline{\prod_{l = 1}^{|\mathcal{D}_{1}|} \overline{\delta_{x_l, y_j}}} & e = (\tilde{y}_{j}, y_{j})
    \end{cases}.
  \end{equation}
\end{defn}

This control clause is similar to the one described in Def. \ref{def:control-clause-wass}, except we no longer have auxiliary vertices for the points in \(\mathcal{D}_{1}\).
The resulting individual mixing unitary operators \(U_{M, e}\) are the same as for the Wasserstein case in Eq. \eqref{eq:mixing-e-was}.
Furthermore, the whole mixing unitary operator \(U_{M}\) is constructed by applying them in sequence as in Eq. \eqref{eq:mixing-unitary-was}.
Note however that since there are fewer edges in this case, there are also fewer individual unitaries.
We apply the operators corresponding to the main vertices first and those corresponding to auxiliary vertices last.
Since this mixing operator also produces some quantum states that don't satisfy the constraints in Eq. \eqref{eq:constraints-dcp}, we follow the same steps as for the Wasserstein distance and we relax the constraints to ensure that the mixing operator preserves the feasibility of the quantum states.

\begin{theorem}\label{thm:feasibility-dcp}
  The mixing operator \(U_M (\beta)\) preserves the feasibility of quantum states.
  That is, if \(\ket{\bm{s}}\) represents a matching that satisfies the constraints
  \begin{equation}\label{eq:relaxed-constraints-dcp}
    \begin{split}
      \sum_{(i,j) \sim v} \delta_{(i,j)} \le 1 \ &, \ \ \forall v \in \mathcal{D}_{1}, \mathcal{D}_{2} \\
      \sum_{e \sim v} \delta_{e} \ge 1 \ &, \ \ \forall v \in \mathcal{D}_{2}
    \end{split}
  \end{equation}
  then \(U_M (\beta) \ket{\bm{s}}\) will also satisfy these constraints.
  Moreover, the minimum of the cost function in Eq. \eqref{eq:cost-dcp} over these constraints is the same as the minimum over the constraints in Eq. \eqref{eq:constraints-dcp}.
\end{theorem}

The mixing operator of Theorem \ref{thm:feasibility-dcp}  produces every solution to the relaxed constraints in Eq. \eqref{eq:relaxed-constraints-dcp} in a single iteration.
Using the mixing operator again changes the amplitudes of the quantum states but it does not create any new states.

\begin{theorem}\label{thm:completeness-dcp}
  Applying each of the \(n \times m + m\) unitaries \(U_{M,e} (\beta)\) in sequence starting with the initial state
  \begin{equation}\label{eq:trivial-dcp}
    \ket*{
      \begin{array}{ccc}
        1 & \cdots & 1 \\
        \vdots & \ddots & \vdots \\
        1 & \cdots & 1 \\
        0 & \cdots & 0
      \end{array}
    }
  \end{equation}
  yields a superposition of all possible matchings that satisfy Eq. \eqref{eq:relaxed-constraints-dcp} with non-zero amplitudes.
\end{theorem}

The proofs of Theorem \ref{thm:completeness-dcp} and \ref{thm:feasibility-dcp} are similar to those of Theorems \ref{thm:feasibility-was} and \ref{thm:completeness-was}, respectively,  so we delegate them to Appendix \ref{sec:proofs}.

\section{Implementations} \label{sec:results}
In this section we provide a few implementations of our algorithm to some simple persistence diagrams.
The first are persistence diagrams of samples from circles which we mostly use to illustrate the quantum algorithm for each distance and show the differences between them.
Next, we add some noise to the samples from these circles, which creates similar but slightly more complicated persistence diagrams, and compare the performance of the two methods.

\subsection{An Illustrative Example}
\label{sec:simple-example}

\begin{figure}[ht]
\centering

  \begin{subfigure}[b]{0.475\textwidth}
  \centering
    \includegraphics[clip, trim = {13 15 12 12}]{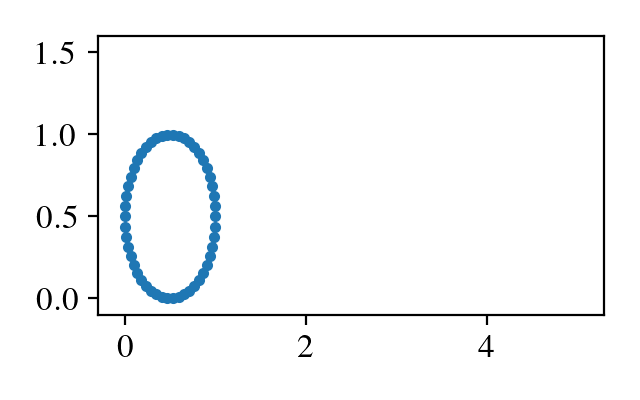}
    \caption{}
    \label{fig:one-circle-cloud}
  \end{subfigure}
  \begin{subfigure}[b]{0.475\textwidth}
  \centering
    \includegraphics[clip, trim = {11 13 9 9}]{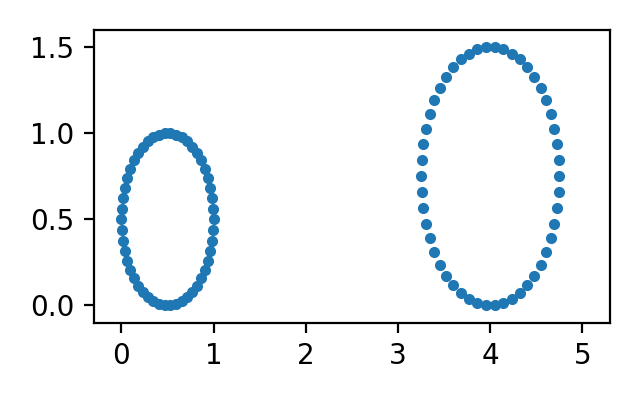}
    \caption{}
    \label{fig:two-circles-cloud}
  \end{subfigure}

  \medskip
  \begin{subfigure}[b]{0.475\textwidth}
  \centering
    \includegraphics[clip, trim = {10 18 12 15}]{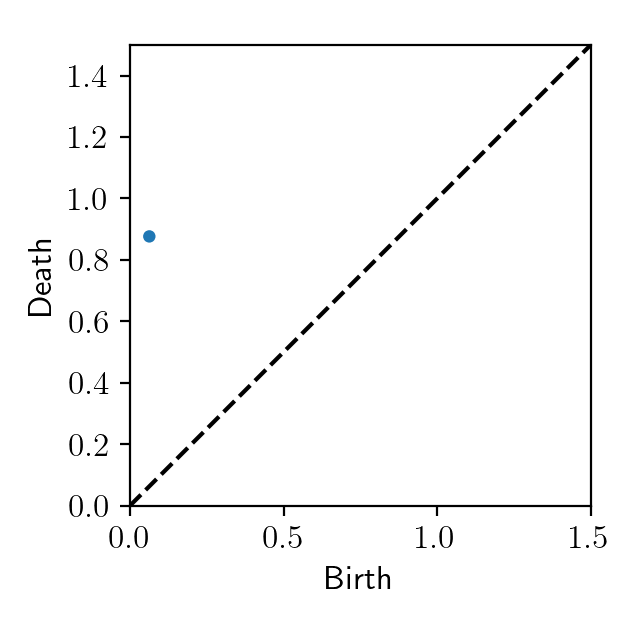}
    \caption{}
    \label{fig:one-circle-diagram}
  \end{subfigure}
  \begin{subfigure}[b]{0.475\textwidth}
  \centering
    \includegraphics[clip, trim = {11 19 11 16}]{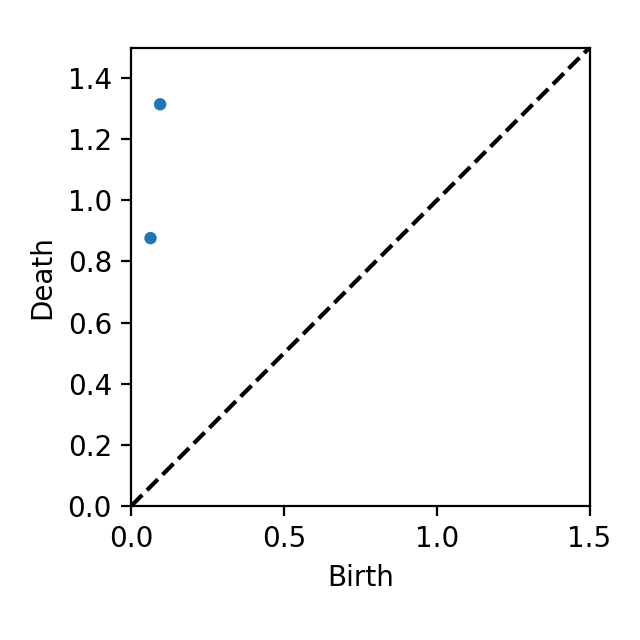}
    \caption{}
    \label{fig:two-circles-diagram}
  \end{subfigure}

  \caption{
    Two point clouds (top), the first (left) sampled from one circle and the second (right) sampled from two circles, along with their respective persistence diagrams (bottom) for dimension 1.
  }
  \label{fig:circles}

\end{figure}

We first use a simple example shown in Fig. \ref{fig:circles} to showcase the corresponding circuits for the Wasserstein and \(d^{c}_{p}\) distances.
One of the data sets consists of points sampled from a circle, the corresponding persistence diagram of dimension 1 has only one point.
The other data set has in addition samples from a second larger circle, and thus the corresponding diagram contains a second point.
So both distances are obtained by matching the shared points \(x_{1}\) and \(y_{1}\) representing the small circle, and penalizing the extra point \(y_{2}\) created by the large circle.
For this example one iteration of the unitaries is more than enough to find the matching that minimizes the cost functions and represents the distance between the diagrams.

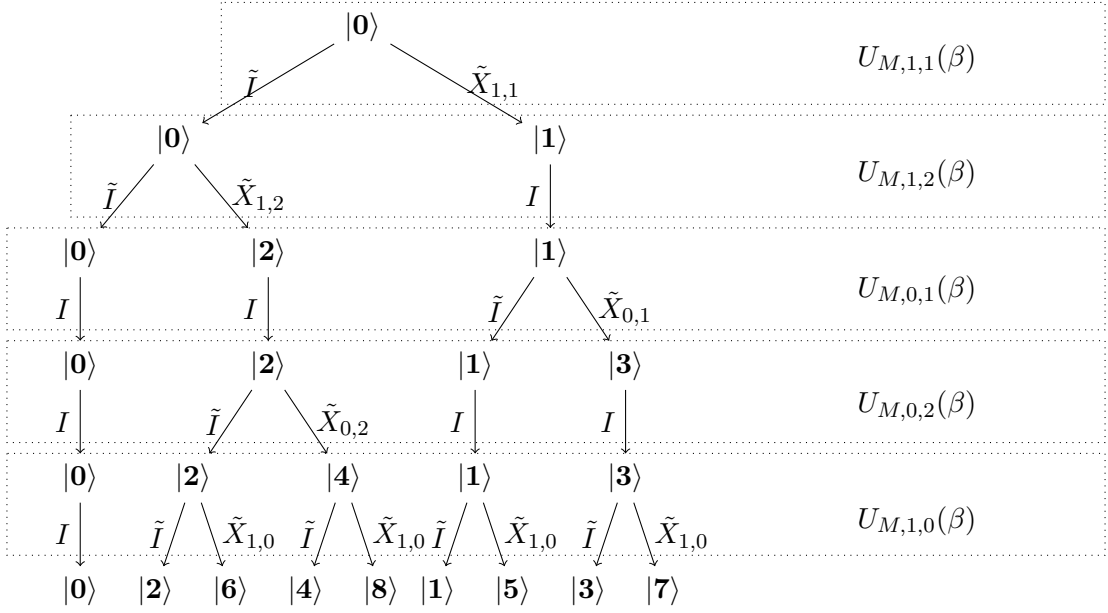
\begin{figure}
  \centering
  \begin{tikzpicture}
    [
    level 1/.style = {sibling distance = 5cm},
    level 2/.style = {sibling distance = 2.5cm},
    level 3/.style = {sibling distance = 2cm},
    level 4/.style = {sibling distance = 2cm},
    level 5/.style = {sibling distance = 1cm},
    ]
    \node (L0) {\(\ket{\bm{0}}\)}
    child { node (L1A) {\(\ket{\bm{0}}\)}
      child { node (L2A) {\(\ket{\bm{0}}\)}
        child { node (L3A) {\(\ket{\bm{0}}\)}
          child { node (L4A) {\(\ket{\bm{0}}\)}
            child { node (L5A) {\(\ket{\bm{0}}\)}
              edge from parent [->] node [left] {\(I\)}
            }
            edge from parent [->] node [left] {\(I\)}
          }
          edge from parent [->] node [left] {\(I\)}
        }
        edge from parent [->] node [left] {\(\tilde{I}\)}
      }
      child { node {\(\ket{\bm{2}}\)}
        child { node {\(\ket{\bm{2}}\)}
          child { node {\(\ket{\bm{2}}\)}
            child { node {\(\ket{\bm{2}}\)}
              edge from parent [->] node [left] {\(\tilde{I}\)}
            }
            child { node {\(\ket{\bm{6}}\)}
              edge from parent [->] node [right] {\(\tilde{X}_{1,0}\)}
            }
            edge from parent [->] node [left] {\(\tilde{I}\)}
          }
          child { node {\(\ket{\bm{4}}\)}
            child { node {\(\ket{\bm{4}}\)}
              edge from parent [->] node [left] {\(\tilde{I}\)}
            }
            child { node {\(\ket{\bm{8}}\)}
              edge from parent [->] node [right] {\(\tilde{X}_{1,0}\)}
            }
            edge from parent [->] node [right] {\(\tilde{X}_{0,2}\)}
          }
          edge from parent [->] node [left] {\(I\)}
        }
        edge from parent [->] node [right] {\(\tilde{X}_{1,2}\)}
      }
      edge from parent [->] node [left] {\(\tilde{I}\)}
    }
    child { node (L1B) {\(\ket{\bm{1}}\)}
      child { node (L2B) {\(\ket{\bm{1}}\)}
        child { node {\(\ket{\bm{1}}\)}
          child { node {\(\ket{\bm{1}}\)}
            child { node {\(\ket{\bm{1}}\)}
              edge from parent [->] node [left] {\(\tilde{I}\)}
            }
            child { node {\(\ket{\bm{5}}\)}
              edge from parent [->] node [right] {\(\tilde{X}_{1,0}\)}
            }
            edge from parent [->] node [left] {\(I\)}
          }
          edge from parent [->] node [left] {\(\tilde{I}\)}
        }
        child { node (L3B) {\(\ket{\bm{3}}\)}
          child { node (L4B) {\(\ket{\bm{3}}\)}
            child { node {\(\ket{\bm{3}}\)}
              edge from parent [->] node [left] {\(\tilde{I}\)}
            }
            child { node (L5B) {\(\ket{\bm{7}}\)}
              edge from parent [->] node [right] {\(\tilde{X}_{1,0}\)}
            }
            edge from parent [->] node [left] {\(I\)}
          }
          edge from parent [->] node [right] {\(\tilde{X}_{0,1}\)}
        }
        edge from parent [->] node [left] {\(I\)}
      }
      edge from parent [->] node [right] {\(\tilde{X}_{1,1}\)}
    }
    ;
    \node (r0) at ($(L0.south east)+(7.0,-0.1)$) {\(U_{M,1,1}(\beta)\)};
    \draw[dotted] ($(L0.north west)+(-1.5,0.0)$)  rectangle ($(L0.south east)+(9.5,-0.7)$);
    \node (r1) at ($(r0.south)+(0.0,-1.2)$) {\(U_{M,1,2}(\beta)\)};
    \draw[dotted] ($(L1A.north west)+(-1.0,0.0)$)  rectangle ($(L1B.south east)+(7.0,-0.7)$);
    \node (r2) at ($(r1.south)+(0.0,-1.2)$) {\(U_{M,0,1}(\beta)\)};
    \draw[dotted] ($(L2A.north west)+(-0.6,0.0)$)  rectangle ($(L2B.south east)+(7.0,-0.7)$);
    \node (r3) at ($(r2.south)+(0.0,-1.2)$) {\(U_{M,0,2}(\beta)\)};
    \draw[dotted] ($(L3A.north west)+(-0.6,0.0)$)  rectangle ($(L3B.south east)+(6.0,-0.7)$);
    \node (r4) at ($(r3.south)+(0.0,-1.2)$) {\(U_{M,1,0}(\beta)\)};
    \draw[dotted] ($(L4A.north west)+(-0.6,0.0)$)  rectangle ($(L4B.south east)+(6.0,-0.7)$);
  \end{tikzpicture}
  \caption{Construction tree for Wasserstein distance of the example in Fig. \ref{fig:circles}, where the quantum states are shown in Table \ref{table:solution-wass-example}. }
  \label{fig:tree-wass}
\end{figure}

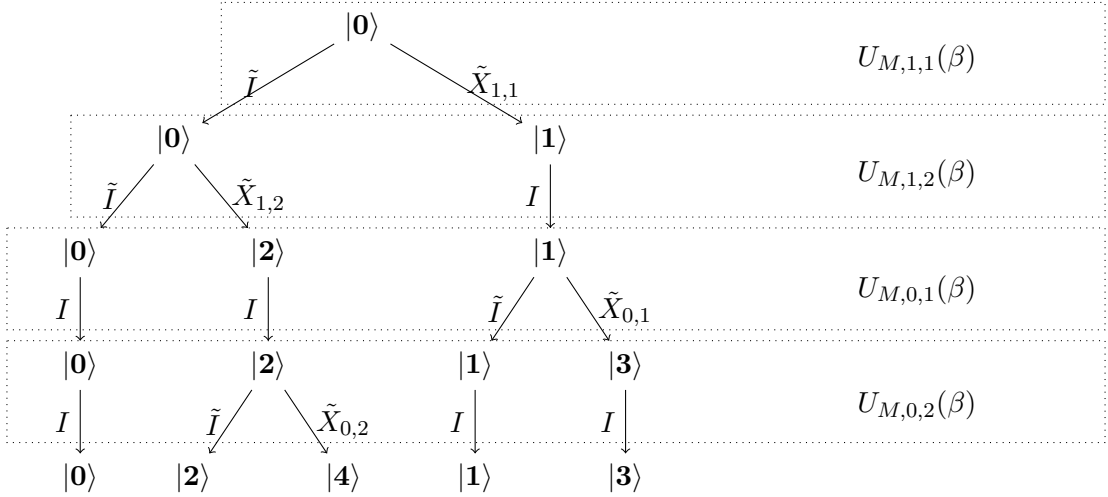
\begin{figure}
  \centering
  \begin{tikzpicture}
    [
    level 1/.style = {sibling distance = 5cm},
    level 2/.style = {sibling distance = 2.5cm},
    level 3/.style = {sibling distance = 2cm},
    level 4/.style = {sibling distance = 2cm},
    ]
    \node (L0) {\(\ket{\bm{0}}\)}
    child { node (L1A) {\(\ket{\bm{0}}\)}
      child { node (L2A) {\(\ket{\bm{0}}\)}
        child { node (L3A) {\(\ket{\bm{0}}\)}
          child { node (L4A) {\(\ket{\bm{0}}\)}
            edge from parent [->] node [left] {\(I\)}
          }
          edge from parent [->] node [left] {\(I\)}
        }
        edge from parent [->] node [left] {\(\tilde{I}\)}
      }
      child { node {\(\ket{\bm{2}}\)}
        child { node {\(\ket{\bm{2}}\)}
          child { node {\(\ket{\bm{2}}\)}
            edge from parent [->] node [left] {\(\tilde{I}\)}
          }
          child { node {\(\ket{\bm{4}}\)}
            edge from parent [->] node [right] {\(\tilde{X}_{0,2}\)}
          }
          edge from parent [->] node [left] {\(I\)}
        }
        edge from parent [->] node [right] {\(\tilde{X}_{1,2}\)}
      }
      edge from parent [->] node [left] {\(\tilde{I}\)}
    }
    child { node (L1B) {\(\ket{\bm{1}}\)}
      child { node (L2B) {\(\ket{\bm{1}}\)}
        child { node {\(\ket{\bm{1}}\)}
          child { node {\(\ket{\bm{1}}\)}
            edge from parent [->] node [left] {\(I\)}
          }
          edge from parent [->] node [left] {\(\tilde{I}\)}
        }
        child { node (L3B) {\(\ket{\bm{3}}\)}
          child { node (L4B) {\(\ket{\bm{3}}\)}
            edge from parent [->] node [left] {\(I\)}
          }
          edge from parent [->] node [right] {\(\tilde{X}_{0,1}\)}
        }
        edge from parent [->] node [left] {\(I\)}
      }
      edge from parent [->] node [right] {\(\tilde{X}_{1,1}\)}
    }
    ;
    \node (r0) at ($(L0.south east)+(7.0,-0.1)$) {\(U_{M,1,1}(\beta)\)};
    \draw[dotted] ($(L0.north west)+(-1.5,0.0)$)  rectangle ($(L0.south east)+(9.5,-0.7)$);
    \node (r1) at ($(r0.south)+(0.0,-1.2)$) {\(U_{M,1,2}(\beta)\)};
    \draw[dotted] ($(L1A.north west)+(-1.0,0.0)$)  rectangle ($(L1B.south east)+(7.0,-0.7)$);
    \node (r2) at ($(r1.south)+(0.0,-1.2)$) {\(U_{M,0,1}(\beta)\)};
    \draw[dotted] ($(L2A.north west)+(-0.6,0.0)$)  rectangle ($(L2B.south east)+(7.0,-0.7)$);
    \node (r3) at ($(r2.south)+(0.0,-1.2)$) {\(U_{M,0,2}(\beta)\)};
    \draw[dotted] ($(L3A.north west)+(-0.6,0.0)$)  rectangle ($(L3B.south east)+(6.0,-0.7)$);
  \end{tikzpicture}
  \caption{Construction tree for the example in Fig. \ref{fig:circles}.}
  \label{fig:tree-dcp}
\end{figure}

We can illustrate how the mixing unitary operator acts on the initial state using a tree in which each level represents the application of an individual mixing unitary to the current state of the system, see Fig. \ref{fig:tree-wass} for the Wasserstein distance and Fig. \ref{fig:tree-dcp} for the \(d^{c}_{p}\) distance.

The quantum states for the Waserstein case are arranged as
\begin{equation}\label{eq:qstates-was-example}
  \ket*{
    \begin{array}{ccc}
      (x_{1}, y_{1}) & (x_{1}, y_{2}) & (x_{1}, \tilde{x}_{1}) \\
      (\tilde{y}_{1}, y_{1}) & (\tilde{y}_{2}, y_{2}) &
    \end{array}
  },
\end{equation}
and the ones encoding possible solutions to the optimizations problem are shown in Table \ref{table:solution-wass-example}.
In particular, the quantum states \(\ket{\bm{0}}, \ket{\bm{7}}, \ket{\bm{8}}\) correspond to matchings that satisfy the original constraints in Eq. \eqref{eq:constraints-was}, while the remaining states \(\ket{\bm{1}}, \dots, \ket{\bm{6}}\) only satisfy the relaxed constraints in Eq. \eqref{eq:relaxed-constraints-was}.

\begin{table}
  \centering
  \caption{Solutions to the Wasserstein optimization problem.}
  \begin{tabular}{ c  c  c }
    \(
    \ket{\bm{0}} \equiv \ket*{
    \begin{array}{ccc}
      1 & 1 & 0  \\
      0 & 0 &
    \end{array}
    } \) &
    \(
    \ket{\bm{1}} \equiv \ket*{
    \begin{array}{ccc}
      0 & 1 & 0  \\
      0 & 0 &
    \end{array}
    } \) &
    \(
    \ket{\bm{2}} \equiv \ket*{
    \begin{array}{ccc}
      1 & 0 & 0  \\
      0 & 0 &
    \end{array}
    } \) \\
    \(
    \ket{\bm{3}} \equiv \ket*{
    \begin{array}{ccc}
      0 & 1 & 0  \\
      1 & 0 &
    \end{array}
    } \) &
    \(
    \ket{\bm{4}} \equiv \ket*{
    \begin{array}{ccc}
      1 & 0 & 0  \\
      0 & 1 &
    \end{array}
    } \) &
    \(
    \ket{\bm{5}} \equiv \ket*{
    \begin{array}{ccc}
      0 & 1 & 1  \\
      0 & 0 &
    \end{array}
    } \) \\
    \(
    \ket{\bm{6}} \equiv \ket*{
    \begin{array}{ccc}
      1 & 0 & 1  \\
      0 & 0 &
    \end{array}
    } \) &
    \(
    \ket{\bm{7}} \equiv \ket*{
    \begin{array}{ccc}
      0 & 1 & 1  \\
      1 & 0 &
    \end{array}
    } \) &
    \(
    \ket{\bm{8}} \equiv \ket*{
    \begin{array}{ccc}
      1 & 0 & 1  \\
      0 & 1 &
    \end{array}
    } \)
  \end{tabular}
  \label{table:solution-wass-example}
\end{table}

The mixing operator consists of a series of individual unitary operators corresponding to each edge.
First, we apply those corresponding to the main edges, in this case \((x_{1}, y_{1})\) and \((x_{1}, y_{2})\).
As shown on the first level of the tree in Fig. \ref{fig:tree-wass}, applying the individual mixing unitary \(U_{M, 1, 1}(\beta)\) defined in Eq. \eqref{eq:mixing-e-was} to the initial state \(\ket{\bm{0}}\) creates the superposition \(\cos(\beta/2) \ket{\bm{0}} - i \sin(\beta/2) \ket{\bm{1}}\), with \(\ket{\bm{0}}, \ket{\bm{1}}\) given in Table \ref{table:solution-wass-example}.
Indeed, one may verify that for the initial state \(\ket{\bm{0}}\) the control clause \(f\) introduced in Def. \ref{def:control-clause-wass} has a value \(f(x_{1}, y_{1}) = 1\).
Next, the individual mixing unitary \(U_{M, 1, 2}\) acts on the superposition generated by the previous step.
Notice that the term involving \(\ket{\bm{1}}\) remains unchanged since the control clause on this state is \(f(x_{1}, y_{2}) = 0\).
However, the control clause is \(f(x_{1}, y_{2}) = 1\) on the state \(\ket{\bm{0}}\), so the corresponding term is split into \(\cos^{2}(\beta/2) \ket{\bm{0}} - i \cos(\beta/2) \sin(\beta/2) \ket{\bm{2}}\).
Thus, after the second individual unitary the state of the system is given by a superposition of \(\ket{\bm{0}}, \ket{\bm{1}}, \ket{\bm{2}}\) as can be seen on the second level in Fig. \ref{fig:tree-wass}.
Note that the initial state corresponds to the solution of the constraints in Eq. \eqref{eq:constraints-was} that matches every point to the diagonal.
Moreover, the individual unitaries for the main edges generate solutions to the relaxed constraints in Eq. \eqref{eq:relaxed-constraints-was} by adding the corresponding main edge to the matching.

The second part of the mixing operator consists of the individual mixing unitaries that correspond to auxiliary edges, that is \((\tilde{y}_{1}, y_{1})\), \((\tilde{y}_{2}, y_{2})\), and \((x_{1}, \tilde{x}_{1})\).
The third and fourth levels in Fig. \ref{fig:tree-wass} show that the individual unitaries \(U_{M, 0, 1}\) and \(U_{M, 0, 2}\) generate quantum states \(\ket{\bm{3}}\) and \(\ket{\bm{4}}\) respectively.
Finally, the individual unitary \(U_{M, 1, 0}\) generates all the remaining solutions \(\ket{\bm{5}}, \ket{\bm{6}}, \ket{\bm{7}}, \ket{\bm{8}}\).
Notice that the unitaries for the auxiliary edges act by removing the corresponding edge from the matching.

On the other hand, the quantum states for the \(d^{c}_{p}\) distance are given by
\begin{equation}
  \label{eq:qstates-dcp-example}
  \ket*{
    \begin{array}{cc}
      (x_{1}, y_{1}) & (x_{1}, y_{2}) \\
      (\tilde{y}_{1}, y_{1}) & (\tilde{y}_{2}, y_{2})
    \end{array}
  },
\end{equation}
and the possible solutions to the optimization problem are depicted in Table \ref{table:solution-dcp-example}.
The quantum states encoding solutions to the original constraints in Eq. \eqref{eq:constraints-dcp} are \(\ket{\bm{0}}, \ket{\bm{3}}, \ket{\bm{4}}\), and those that represent solutions of the relaxed constraints in Eq. \eqref{eq:relaxed-constraints-dcp} are \(\ket{\bm{1}}, \ket{\bm{2}}\).

\begin{table}
  \centering
  \caption{Solutions to the alternative optimization problem.}
  \begin{tabular}{ c  c  c }
    \(
    \ket{\bm{0}} \equiv \ket*{
      \begin{array}{cc}
        1 & 1 \\
        0 & 0
      \end{array}
    } \) &
    \(
    \ket{\bm{1}} \equiv \ket*{
      \begin{array}{cc}
        0 & 1 \\
        0 & 0
      \end{array}
    } \) &
    \(
    \ket{\bm{2}} \equiv \ket*{
      \begin{array}{cc}
        1 & 0 \\
        0 & 0
      \end{array}
    } \) \\
    \(
    \ket{\bm{3}} \equiv \ket*{
      \begin{array}{cc}
        0 & 1 \\
        1 & 0
      \end{array}
    } \) &
    \(
    \ket{\bm{4}} \equiv \ket*{
      \begin{array}{cc}
        1 & 0 \\
        0 & 1
      \end{array}
    } \) &
  \end{tabular}
  \label{table:solution-dcp-example}
\end{table}

Similar to the Wasserstein case, the initial state \(\ket{\bm{0}}\) represents penalizing all points with the constant \(c\).
The first and second level of the tree in Fig. \ref{fig:tree-dcp} show that the individual mixing unitaries corresponding to main edges \(U_{M, 1, 1}\) and \(U_{M, 1, 2}\) generate the quantum states \(\ket{\bm{1}}\) and \(\ket{\bm{2}}\) respectively by adding main edges to the matching.
Then, the last two levels in Fig. \ref{fig:tree-dcp} show how individual operators corresponding to auxiliary edges \(U_{M, 0, 1}\) and \(U_{M, 0, 2}\) remove auxiliary edges to generate the states \(\ket{\bm{3}}\) and \(\ket{\bm{4}}\) respectively.
Note that in this case the edge \(x_{1}, \tilde{x}_{1}\) is not present, which results in one less individual unitary and significantly fewer possible solutions.
The quantum circuits for the mixing operator of both the Wasserstein and \(d^{c}_{p}\) distances can be found in Appendix \ref{sec:quantum-circuits}.

\subsection{Circles vs Noisy Circles}

\begin{figure}[ht]
\centering

  \begin{subfigure}[b]{0.475\textwidth}
  \centering
    \includegraphics[clip, trim = {11 13 10 10}]{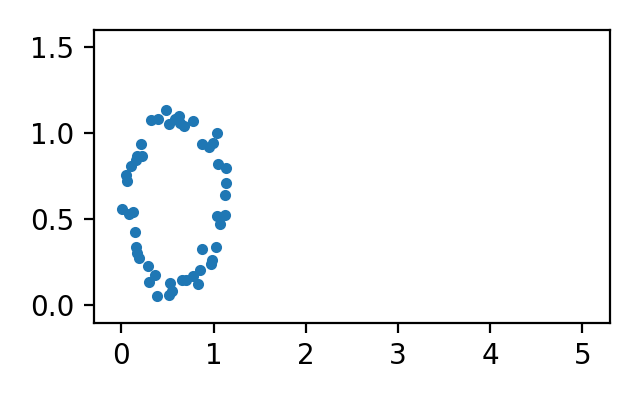}
    \caption{}
    \label{fig:noisy-one-circle-cloud}
  \end{subfigure}
  \begin{subfigure}[b]{0.475\textwidth}
  \centering
    \includegraphics[clip, trim = {11 13 10 10}]{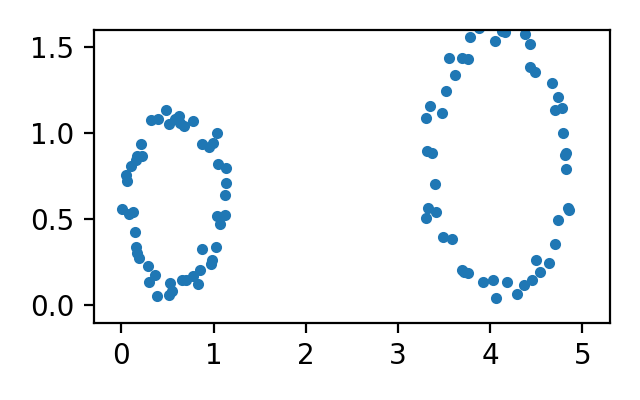}
    \caption{}
    \label{fig:noisy-two-circles-cloud}
  \end{subfigure}

  \medskip
  \begin{subfigure}[b]{0.475\textwidth}
  \centering
    \includegraphics[clip, trim = {10 19 11 17}]{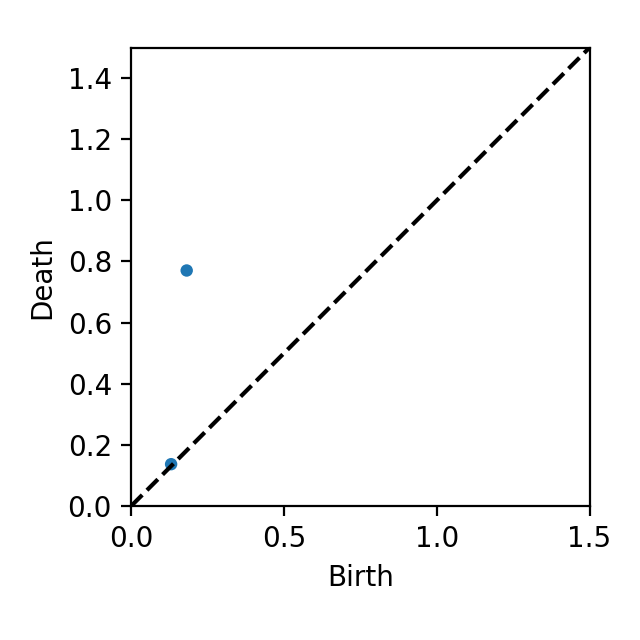}
    \caption{}
    \label{fig:noisy-one-circle-diagram}
  \end{subfigure}
  \begin{subfigure}[b]{0.475\textwidth}
  \centering
    \includegraphics[clip, trim = {11 19 11 17}]{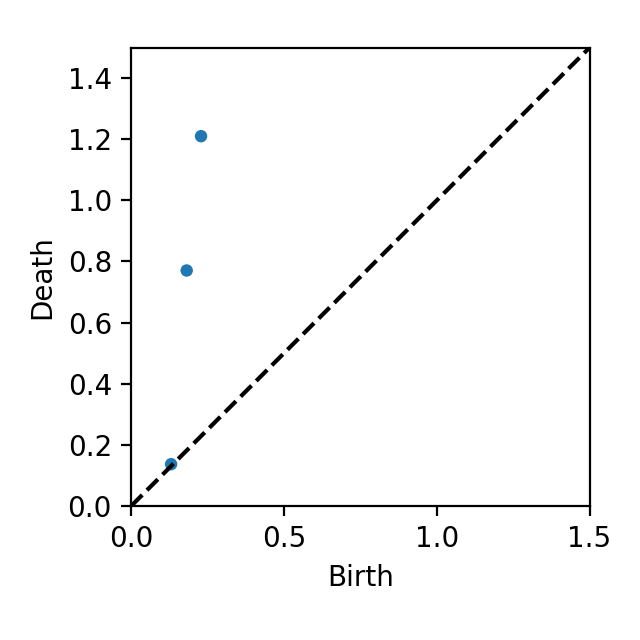}
    \caption{}
    \label{fig:noisy-two-circles-diagram}
  \end{subfigure}

  \caption{
    Two point clouds (top) resulting from adding noise to the data sets shown in Fig. \ref{fig:one-circle-cloud} and \ref{fig:two-circles-cloud}, along with their respective persistence diagrams (bottom) for dimension 1.
  }
  \label{fig:noisy-circles}

\end{figure}

Next, we added noise to the samples from the circles in Fig. \ref{fig:one-circle-cloud} and \ref{fig:two-circles-cloud} to create slightly more complex data sets (shown in Fig. \ref{fig:noisy-one-circle-cloud} and \ref{fig:noisy-two-circles-cloud}).
Their corresponding persistence diagrams of dimension one have an extra point each generated by the noise, see Fig. \ref{fig:noisy-one-circle-diagram} and \ref{fig:noisy-two-circles-diagram}.

\begin{figure}[ht]
\centering

  \begin{subfigure}[b]{0.475\textwidth}
  \centering
    \includegraphics[clip, trim = {12 12 10 10}]{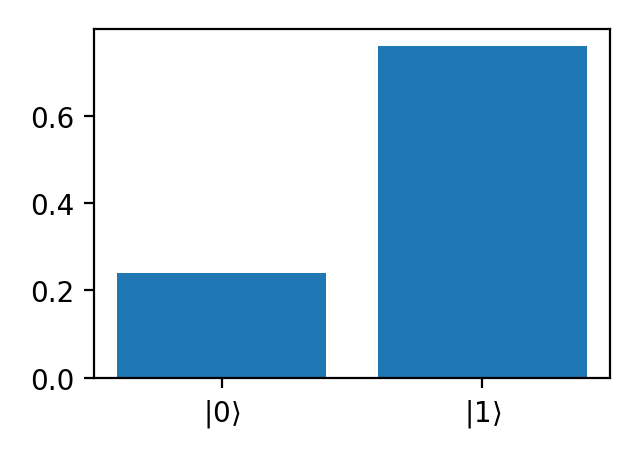}
    \caption{}
    \label{fig:distribution-wass-clean}
  \end{subfigure}
  \begin{subfigure}[b]{0.475\textwidth}
  \centering
    \includegraphics[clip, trim = {12 12 10 10}]{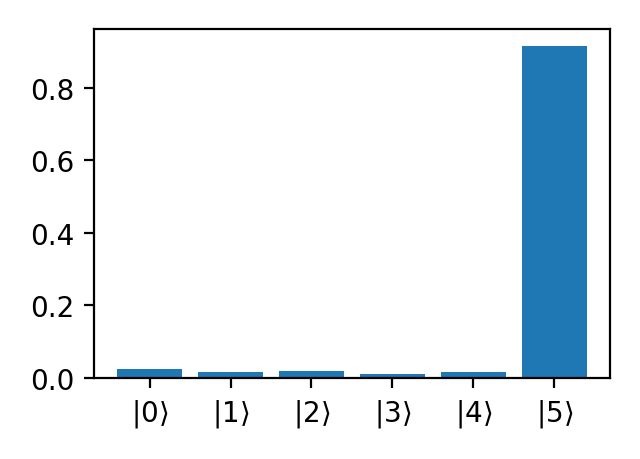}
    \caption{}
    \label{fig:distribution-wass-noise}
  \end{subfigure}

  \medskip
  \begin{subfigure}[b]{0.475\textwidth}
  \centering
    \includegraphics[clip, trim = {12 12 10 10}]{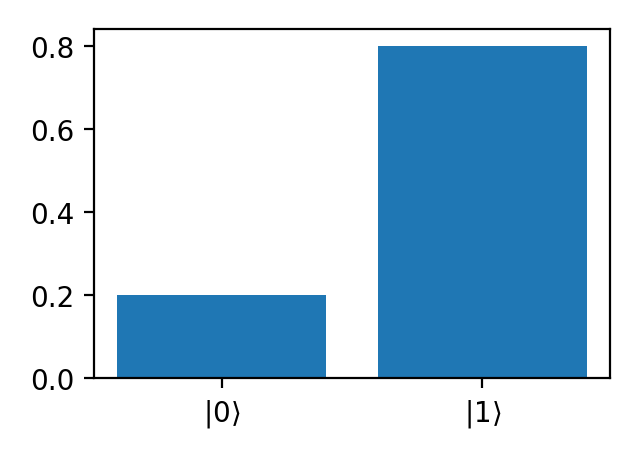}
    \caption{}
    \label{fig:distribution-dcp-clean}
  \end{subfigure}
  \begin{subfigure}[b]{0.475\textwidth}
  \centering
    \includegraphics[clip, trim = {12 12 10 10}]{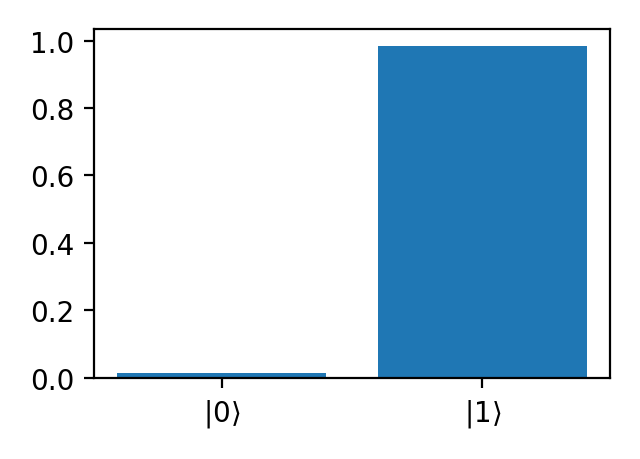}
    \caption{}
    \label{fig:distribution-dcp-noise}
  \end{subfigure}

  \caption{
    Frequency of measurements after one iteration of our QAOA algorithm for the Wasserstein (top) and \(d^{c}_{p}\) (bottom) distances on the data sets from Fig. \ref{fig:circles} (left) and Fig. \ref{fig:noisy-circles} (right).
  }
  \label{fig:measurements}

\end{figure}

We used the Qiskit library for Python to simulate our quantum algorithms on the noisy and noiseless data sets.
Only one iteration of the mixing and problem operators was used in the simulations.
The rotation angles were obtained by minimizing the cost functions in Eq. \eqref{eq:cost-was} and \eqref{eq:cost-dcp}.
The results of the simulations using the optimal angles can be found in Fig. \ref{fig:measurements}.
Note that the parameters from Definitions \ref{def:was} and \ref{def:dcp} chosen for all simulations are \(q = \infty\), \(p = 2\), and \(c = 0.2\).

Figure \ref{fig:distribution-wass-clean} shows the observed quantum states for the Wasserstein distance on the noiseless data sets from Fig. \ref{fig:circles}.
The first observed state \(\ket{0}\) corresponds to
\(\ket*{
  \begin{array}{ccc}
    1 & 0 & 1 \\
    0 & 1 &
  \end{array}
}\)
which matches the circle in Fig. \ref{fig:one-circle-cloud} to the large circle in Fig. \ref{fig:two-circles-cloud}.
On the other hand, the state observed with the most frequency (\(\ket{1}\)) corresponds to
\(\ket*{
  \begin{array}{ccc}
    0 & 1 & 1 \\
    1 & 0 &
  \end{array}
}\),
the matching that yields the Wasserstein distance.

Similarly, the results for the \(d^{c}_{p}\) distance on the noiseless data sets are depicted in Fig. \ref{fig:distribution-dcp-clean}.
As in the case of the Wasserstein distance, the observed quantum states are \(\ket{0}\) corresponding to
\(\ket*{
  \begin{array}{cc}
    1 & 0 \\
    0 & 1
  \end{array}
}\),
and the one with the highest frequency, \(\ket{1}\), equivalent to the matching that yields the \(d^{c}_{p}\) distance,
\(\ket*{
  \begin{array}{cc}
    0 & 1 \\
    1 & 0
  \end{array}
}\).

For the noisy data sets from Fig. \ref{fig:noisy-circles} one iteration of the quantum algorithm was not good enough for the Wasserstein distance as shown in Fig. \ref{fig:distribution-wass-noise}.
This time the most observed state \(\ket{5}\) corresponds to the initial state
\(\ket*{
  \begin{array}{cccc}
    1 & 1 & 1 & 0 \\
    1 & 1 & 1 & 0 \\
    0 & 0 & 0 &
  \end{array}
}\)
which is definitely not the matching that yields the Wasserstein distance.
The other observed states are solutions to the relaxed constraints in Eq. \eqref{eq:relaxed-constraints-was} but not the original constraints in Eq. \eqref{eq:constraints-was}, and therefore also do not yield the Wasserstein distance.
In contrast, the algorithm for the \(d^{c}_{p}\) distance did find the solution for the noisy data sets, as illustrated in Fig. \ref{fig:distribution-dcp-noise}.
The least observed quantum state \(\ket{0}\) corresponds to the solution that matches the holes created by noise as well the circle in Fig. \ref{fig:noisy-one-circle-cloud} to the large circle in Fig. \ref{fig:noisy-two-circles-cloud},
\(\ket*{
  \begin{array}{ccc}
    0 & 1 & 1 \\
    1 & 1 & 0 \\
    1 & 0 & 1
  \end{array}
}\).
The state with the highest frequency, \(\ket{1}\), is equivalent to the matching that yields the \(d^{c}_{p}\) distance,
\(\ket*{
  \begin{array}{ccc}
    0 & 1 & 1 \\
    1 & 0 & 1 \\
    1 & 1 & 0
  \end{array}
}\).
This suggests that computing the \(d^{c}_{p}\) distance can be more efficient since the circuit for the Wasserstein distance already requires more iterations for such a simple example.

\section{Conclusion and Discussion} \label{sec:conclusion}
The unitary operator for the cost Hamiltonian requires the application of single qubit rotation operators \(R_{Z}\), \(n \times m + n + m\) in the case of the Wasserstein distance and \(n \times m + m\) for the \(d^{c}_{p}\) distance, all of which can be performed in parallel.
On the other hand, the mixing operator needs to perform single qubit rotations \(R_{X}\) with a control clause \(f\) for each edge in succession.
This step requires \(\mathcal{O}(n m)\) operations since there are \(n \times m + n + m\) edges in the Wasserstein graph and \(n \times m + m\) in the \(d^{c}_{p}\) graph.
Classical methods to compute the distance between persistence diagrams use the Hungarian algorithm that has a cost of \(\mathcal{O}(n^{2} m)\) operations, so our quantum algorithm could provide an advantage.

As was briefly discussed in Section \ref{sec:background} one can use geometric properties of the persistence diagrams to reduce the number of matchings that are considered feasible.
In the case of our quantum algorithm this could be implemented as simple inequality tests during the process of computing the pairwise distances between points, then any edges that fail the test can be ignores during the algorithm therefore reducing the number of qubits and individual mixing operators.

While both distances result in the same overall complexity, the quantum algorithm for the \(d^{c}_{p}\) distance uses fewer qubits and therefore needs fewer mixing operations and controls.
This could provide an advantage over the Wasserstein distance in the NISQ era because of the need to implement error mitigation techniques, and the limited gate sets and connectivity between qubits in some of the available machines.

Some concerns have been raised about the performance of QAOA applied to MaxCut problems on bipartite graphs \cite{farhi2020quantum}, in particular if the distance between vertices is large then the algorithm needs several iterations in order to obtain a high approximation ratio.
Fortunately we don't believe this to be an issue for our algorithm since the graphs it considers are a combination of a complete bipartite graph, in which the maximum distance between vertices is 2, and the auxiliary vertices.

\acknowledgments

This work has been partially supported by the NSF  DMS-2012609, and DGE-2152168.

\bibliographystyle{unsrt}
\bibliography{references}

\begin{appendix}

\section{Proofs of Theorems \ref{thm:feasibility-dcp} and \ref{thm:completeness-dcp}} \label{sec:proofs}

\textbf{Proof of Feasibility: Theorem \ref{thm:feasibility-dcp}.}
We first give some intuition about these new constraints and show that the minimum remains the same.
Notice that the first constraint asks that a point \(v\) in one of the persistence diagrams is matched at most once to points in the other diagram.
The second constraint requires that every point in the largest persistence diagram is matched at least once.
This means that points in \(\mathcal{D}_{2}\) could be matched more than once, but only one of the edges will go towards a point in the other diagram and any other edges connect to auxiliary vertices.
So the solutions to these constraints will be matchings that represent a one-to-one function \(\phi : \mathcal{D}_{1} \to \mathcal{D}_{2}\) or matchings that result of adding connections to auxiliary vertices.
In particular, any solution to the constraints in Eq. \eqref{eq:constraints-dcp} is also a solution to Eq. \eqref{eq:relaxed-constraints-dcp}, and any solution to the latter that is not also a solution to the former must have a greater cost.

Now, given a quantum state \(\ket{s}\) which satisfies the constraints in Eq. \eqref{eq:relaxed-constraints-dcp}, we show that \(U_{M,e}(\beta) \ket{s}\) also satisfies the constraints for arbitrary \(e\in E\).
This in turn proves that \(U_M(\beta)\) preserves the feasibility of quantum states as it is defined as a product of the individual unitaries.

From Eq. \eqref{eq:mixing-e-was} note that there are only two possible cases for \(U_{M,e}(\beta) \ket{s}\) depending on the value of \(f(e)\).
First, if \(f(e) = 0\) the output state is simply \(\ket{s}\).
On the other hand, when \(f(e) = 1\) the resulting state is \(\cos\beta I \ket{s} - i \sin\beta X_{e} \ket{s}\).
The first term is just \(\ket{s}\) scaled by a constant, so we only need to verify that the second term is feasible.

Notice that the second term \(X_{e}\) flips the qubit \(|e\ra\) but it is only applied when \(f(e) = 1\).
If \(e = (x_{i},y_{j})\) is an edge between points in the diagrams, \(f(e) = 1\) if and only if \(\delta_{e}, \delta_{i,k}, \delta_{l,j} = 0\) for all \(l \ne i\) and \(k \ne j\), that is, the edge \(e = (x_{i}, y_{j})\) is added to the matching if there are no other main edges containing vertices \(x_{i}\) or \(y_{j}\).
Since this operation adds an edge, the second constraint in Eq. \eqref{eq:relaxed-constraints-dcp} is trivially satisfied, moreover the first constraint is satisfied because the edge is only added if that sum is equal to zero.
On the other hand, if \((\tilde{y}_{j}, y_{j})\) is an edge to an auxiliary vertex, \(f(e) = 1\) whenever \(\delta_{e} = 1\) and at least one of \(\delta_{l,j}\) for \(1 \le l \le |\mathcal{D}_{1}|\) is non-zero, in other words, we remove the edge \((\tilde{y}_{j}, y_{j})\) from the matching if there is a main edge connecting the corresponding vertex \(y_{j}\) to a vertex in the other diagram.
So, the first constraint in Eq. \eqref{eq:relaxed-constraints-dcp} is unaffected and the second one is still satisfied as we only remove an edge if the corresponding sum is at least 2.

All in all, \(U_{M,e}(\beta) \ket{s}\) is a superposition of quantum states whose corresponding matchings satisfy the constraints in Eq. \eqref{eq:relaxed-constraints-dcp}.

\textbf{Proof of Completeness: Theorem \ref{thm:completeness-dcp}.}
Notice that one may obtain all solutions to Eq. \eqref{eq:relaxed-constraints-dcp} by finding every possible choice for the main \(n \times m\) edges that satisfy the first constraint, and then for each of these obtaining the different combinations of the \(m\) auxiliary edges that satisfy the second constraint.
So, we first apply the unitaries corresponding to the main edges to produce the solutions to the first constraint while keeping all the auxiliary edges on.
After which we use the remaining unitaries to get the combinations of auxiliary edges.
Since every unitary keeps the quantum state it acts on, either whole or scaled by \(\cos\beta\), this process yields all matchings that satisfy Eq. \eqref{eq:relaxed-constraints-dcp}.

Let \(e_1, \dots, e_N\) be any ordering of the \(n \times m\) main edges, we want to prove that applying their corresponding unitaries in sequence will produce a superposition of all solutions to the first constraint in Eq. \eqref{eq:relaxed-constraints-dcp}.
We proceed by induction, where we will prove that applying \(U_{M,e_k}\) to the superposition of all feasible matchings using only edges \(e_1, \dots, e_{k-1}\), yields a superposition of all feasible matchings using only edges \(e_1, \dots, e_{k}\).
Since the case for \(k = 1\) starts with the quantum state in Eq. \eqref{eq:trivial-dcp}, \(f(e_1)\) will always equal 1 and the result of applying \(U_{M,e_1}\) is a superposition of this trivial matching and the matching which includes only edge \(e_1\).
For \(k > 1\), assume we have a superposition of all feasible matchings with edges \(e_1, \dots, e_{k-1}\) and notice that applying \(U_{M,e_k}\) to any one of these matchings can have two possible outcomes.
Indeed if \(f(e_k) = 0\) the matching remains unchanged.
On the other hand, if \(f(e_k) = 1\) the output will be a combination of the input matching and the matching that results from adding edge \(e_k\).
Therefore, \(U_{M, e_k}\) will retain all the matchings with only edges \(e_1, \dots, e_{k-1}\) and produce all feasible matchings that result from adding edge \(e_k\), which yields a superposition over all feasible matchings that use only edges \(e_1, \dots, e_k\).

Now let \(e_{1}', \dots, e_{m}'\) be any ordering of the \(m\) auxiliary edges.
We prove in a similar way that given the superposition that results of the previous step, applying the auxiliary unitaries in sequence yields a superposition of all possible solutions to Eq. \eqref{eq:relaxed-constraints-dcp}.
Indeed, applying \(U_{M, e_{1}'}\) to each of the matchings in the superposition from the previous step gives either the same matching when \(f(e_{1}') = 0\) or a combination of the input matching and the one that results of removing edge \(e_{1}'\) when \(f(e_{1}') = 1\), producing all feasible choices for \(e_{1}'\).
In the same manner, applying \(U_{M, e_{k}'}\) to the superposition with all feasible choices for edges \(e_{1}', \dots, e_{k-1}'\) will produce a superposition with all feasible combinations for edges \(e_{1}', \dots, e_{k}'\).

\section{Quantum Circuits} \label{sec:quantum-circuits}

\begin{figure}[btp]
  \centering
  \includegraphics[clip, trim = {0 0 0 0}, width = \linewidth]{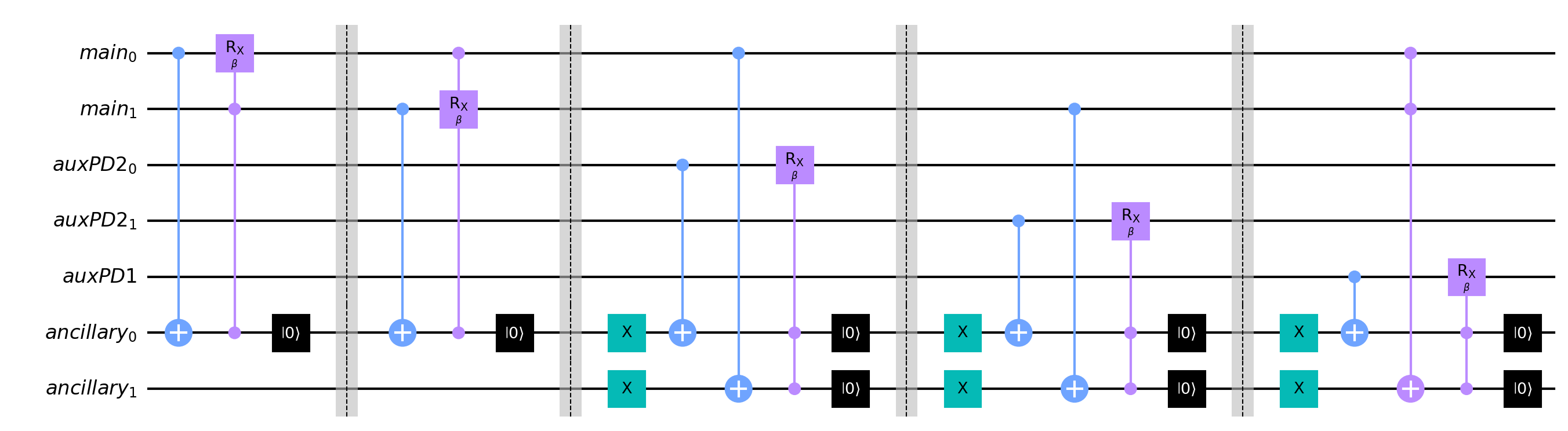}
  \captionof{figure}{
    Quantum circuit for the mixing unitary operator of the wasserstein distance between the persistence diagrams shown in Fig. \ref{fig:circles}.
  }
  \label{fig:circuit-wass}
\end{figure}

Fig. \ref{fig:circuit-wass} shows the quantum circuit for the mixing unitary operator corresponding to the Wasserstein distance of the example introduced in Section \ref{sec:simple-example}.
Barriers separate the different individual mixing operators, and the qubits are arranged in a main register for the main edges between points in the diagrams, auxiliary registers for the edges to auxiliary vertices, and the ancillas needed to compute control clauses.

\begin{figure}[btp]
  \centering
  \includegraphics[clip, trim = {0 0 0 0}, width = \linewidth]{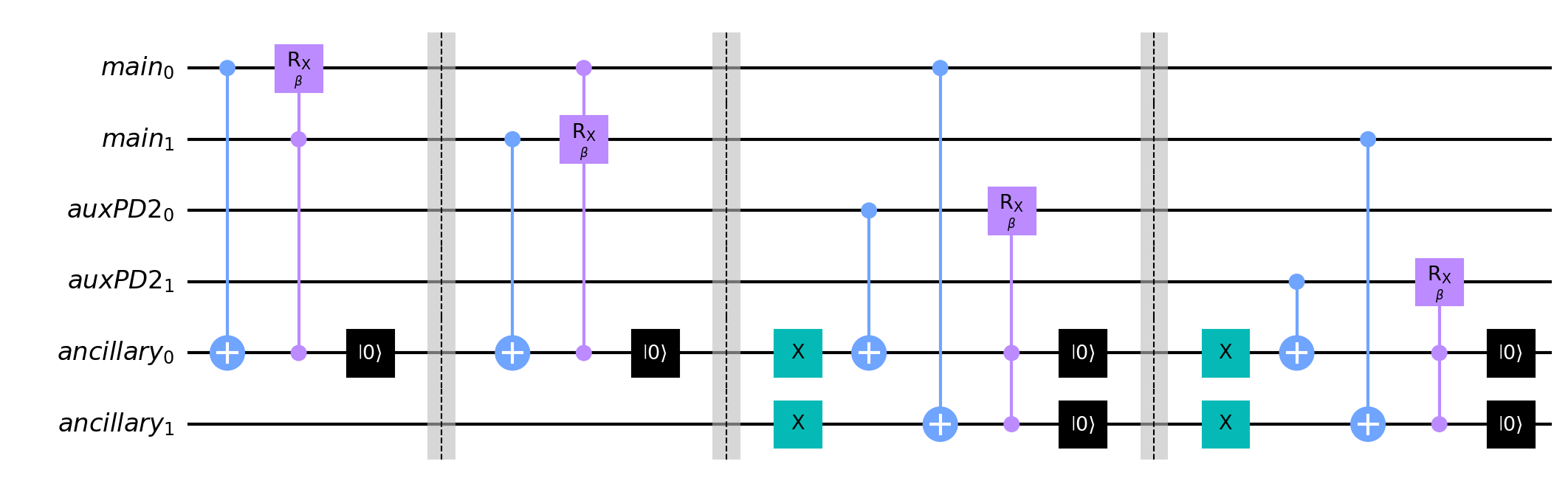}
  \captionof{figure}{
    Quantum circuit for the mixing unitary operator of the \(d^{c}_{p}\) distance between the persistence diagrams shown in Fig. \ref{fig:circles}.
  }
  \label{fig:circuit-dcp}
\end{figure}

On the other hand, Fig. \ref{fig:circuit-dcp} displays the quantum circuit for the mixing unitary operator of the \(d^{c}_{p}\) distance.
Notice that the last individual mixing unitary is not present in this case, which illustrates the possible advantage of using this distance since the individual mixing unitaries are by far the most complex part of the algorithm.
Control clauses used to construct the mixing operator are the only sections of our algorithm that involve interactions between several qubits, thus removing some of these sections could provide an advantage when the connectivity between qubits is limited.
Of course, as mentioned earlier the number of qubits and gates needed to compute the \(d^{c}_{p}\) distance is less than for the Wasserstein distance.

\end{appendix}

\end{document}